\documentclass [12pt,preprint]{aastex}
\usepackage{epsfig}
\begin{document}
\voffset-1cm
\newcommand{\gsim}{\hbox{\rlap{$^>$}$_\sim$}}
\newcommand{\lsim}{\hbox{\rlap{$^<$}$_\sim$}}

\title{The rapid decline of the prompt emission in
       Gamma-Ray Bursts}

\author{Shlomo Dado\altaffilmark{1}, Arnon Dar\altaffilmark{1} and A. De
R\'ujula\altaffilmark{2}}

\altaffiltext{1}{dado@phep3.technion.ac.il, arnon@physics.technion.ac.il,
dar@cern.ch.\\
Physics Department and Space Research Institute, Technion, Haifa 32000,
Israel.}
\altaffiltext{2}{alvaro.derujula@cern.ch; Theory Unit, CERN,
1211 Geneva 23, Switzerland. \\
Physics Department, Boston University, USA.}

\begin{abstract}

Many gamma ray bursts (GRBs) have been observed with the Burst-Alert and 
X-Ray telescopes of the Swift satellite. The successive `pulses' of these 
GRBs end with a fast decline and a fast spectral softening, until they are 
overtaken by another pulse, or the last pulse's decline is overtaken by a 
less rapidly-varying `afterglow'. The fast decline-phase has been 
attributed, in the currently-explored standard fireball model of GRBs, to 
`high-latitude' synchrotron emission from a collision of two conical 
shells. This high latitude emission does not explain the observed 
spectral softening. In contrast, the temporal behaviour and the spectral 
evolution during the fast-decline phase agree with the predictions of the 
cannonball model of GRBs.

\end{abstract}


\section{Introduction}

Since the launch of the Swift spacecraft, precise data from its Burst 
Alert Telescope (BAT) and X-Ray Telescope (XRT) have been obtained on the 
spectral and temporal behaviour of the X-ray emission in $\gamma$-ray 
bursts (GRBs) and X-ray flashes (XRFs). These data have already been used 
to test the most-studied theories of long duration GRBs and their 
afterglows (AGs), the {\it Fireball} (FB) models (see, e.g., Piran 1999, 
2000, 2005; Zhang \& M\'esz\'aros~2004; M\'esz\'aros~2002, 2006,
Zhang~2007; and 
references therein) and the {\it Cannonball} (CB) model [see, e.g.,~Dar \& 
De R\'ujula~2004 (hereafter DD2004); Dado, Dar \& De R\'ujula~2002, 2003, 
2007a, 20007b (hereafter DDD2002, DDD2003, DDD2007a, DDD2007b), and 
references therein].

The general behaviour of the Swift X-ray light curves has been described 
as `canonical' (Nousek et al.~2006; O'Brien et al.~2006; Zhang et 
al.~2007), and is illustrated in Figs.~\ref{f1}a,b, \ref{f2}a for XRF 060218, GRB 060904a 
and GRB 061121.  When measured early enough, the X-ray emission has peaks 
that coincide with the $\gamma$-ray peaks of the GRB. The prompt emission 
has a fast decline after the last detectable peak of the GRB. In most 
cases, the rapid decline ends within a couple of hundreds of seconds. 
Thereafter, it turns into a much flatter `plateau', typically lasting 
thousands to tens of thousands of seconds. Finally, the X-ray light curve, 
within a time order of one day, steepens into a power-law decline which 
lasts until the X-ray AG becomes too dim to be detected. Often, there are 
also X-ray peaks during the fast-decline phase or even later, not 
coinciding with a detectable $\gamma$-ray activity. There is a continuous
transition of X-ray light curve shapes from the `canonical' ones to the
ones that are well described by a single-power decay, e.g.~GRB 061126,
see Fig.~\ref{f2}b.

Neither the general trend, nor the frequently complex structure of the 
Swift X-ray data were correctly predicted by (or can be easily 
accommodated within) the standard FB models (see, e.g., 
Piran~1999, 2000, 2005; ~Zhang \& M\'esz\'aros 2004, for 
reviews of the pre-Swift standard FB model, and Kumar et al.~2007; 
Burrows \& Racusin~2007; Kocevski \& Butler~2007: Urata et al.~2007; 
Zhang, Liang \& Zhang~2007; Yonetoku et al.~2007; Liang et al.~2007
for recent comparisons with Swift data). 

The situation in the CB model (Dado, Dar \& De R\'ujula~2004, hereafter 
DDD2004) is different. The model offered a good description, based on a 
specific synchrotron-radiation (SR) mechanism, of the AGs of all 
`classical' GRBs (DDD2002, DDD2003) of known redshift, and allowed one to 
extract the relevant parameters of the CBs of GRBs and XRFs. The 
consequent predictions for the `prompt' $\gamma$-rays, based on an 
explicit inverse Compton scattering (ICS) mechanism, were simple and 
successful (DD2004). As shown in DDD2007a,b (and references therein) for 
`Swift-era' data, the CB model, with no modification, correctly predicts 
the temporal and spectral behaviour of the prompt and AG phases.

In this paper we confront the Swift's observations with the predictions of 
the FB and CB models for the spectral evolution during the fast-decline 
phase of the prompt emission. In the FB model this phase was interpreted 
(e.g., M\'esz\'aros~2006; Liang et al.~2006; O'Brien et al.~2006; Yamazaki 
et al.~2006) as the `curvature effect' or `high-latitude' emission of 
colliding shells (Fenimore et al.~1996; Kumar \& Panaitescu~2000; 
Dermer~2004). Relative to photons centrally emitted on the line of sight 
to an on-axis observer, photons from off-axis latitudes arrive later and 
with smaller number density and energy. The consequent spectral behaviour 
is entirely different from that observed (see, e.g., Zhang et al.~2007). 
In the CB model the properties of the fast-declining phase are also 
dominantly `geometrical'. A GRB's $\gamma$-ray pulses and their sister 
X-ray flares are made by ICS of light in a `glory reservoir' bathing the 
circumburst material (DD2004). This light becomes, in a very specific 
manner, less abundant and more radially-directed with distance from the 
parent star. These simple facts result in the correct description of the 
temporal behaviour and spectral evolution of GRBs, before, during, and 
after the fast-decline phase.

In the CB model it is possible in principle to fit the spectral energy 
flux of a GRB in a given energy band, as a function of time, and determine 
the parameters partaking in a complete prediction of the spectrum at any 
time in the fit interval. But the public Swift spectral data is limited to 
a `hardness ratio' between the counting rates in the 1.5-10 keV and 
0.3-1.5 keV bands (Evans et al.~2007). To convert these rates into a more 
explicit spectral information one must correct for instrumental 
efficiency, subtract the background and correct for X-ray absorption in 
the host galaxy, in the IGM, and in our Galaxy. 
The unabsorbed spectra as functions of time is not generally available. 
However, the unabsorbed spectral energy flux in the X-ray band, 
parametrized as  $F_\nu\! \propto\! \nu^{-\beta}\, t^{-\alpha}$, is available
in the form of the 
fitted time-dependent power-law spectral index, $\beta(t)$, for a set
 of X-ray light curves measured with Swift's XRT 
(Zhang et al.~2007 and {\it http://swift.physics.unlv.edu/xrt.html}). Such a 
parametrization is not a faithful description of an exponentially cutoff 
power-law, a Band function, or the spectrum predicted by the CB 
model,  similar to the Band function for typical parameters (DD2004). 
Moreover, to extract $\beta(t)$, the data from 
different time intervals was coadded, smoothing the time-dependence 
of the {\it effective} fitted photon index. This can be seen by comparing 
the effective indices $\Gamma(t)$ in $dN_\gamma/dE\!\propto\!E^{-\Gamma}$,
reported in the cited web-page, with
the hardness ratios reported  
in the Swift light-curve repository (Evans et al.~2007).
The spectral indices $\beta$
and $\Gamma$ are related by $\beta\! =\! \Gamma\! -\! 1$.

We do not have all of the information needed for a decisive comparison 
between the spectral behaviour during the fast decline phase of the prompt 
emission. But the variation with time of the hardness ratio and of the 
effective spectral index during the fast decline phase are so spectacular 
and well correlated to the light curves, that an approximate analysis 
suffices to prove our points. We demonstrate this for the hardness ratios of five
Swift GRBs with well sampled X-ray light curves during the fast decline 
phase and for fourteen other GRBs with an extracted effective time-dependent
spectral index.

\section{High-latitude emission in the fireball model} 

In the FB model, GRB pulses are produced by synchrotron radiation emitted 
by a shock-accelerated electrons, following collisions between conical 
shells ejected by a central engine (Rees \& M\'esz\'aros~1994, see Zhang et 
al.~2007 for detailed discussion). Consider a spherical shell, arbitrarily 
thin, that expands with a Lorentz factor 
$\gamma\!\equiv\!1/\sqrt{1\!-\!\,(v/c)^2}$. 
Assume that when two shells collide at a radius 
$R$, all points emit isotropically in their rest frame an 
arbitrarily short pulse of radiation. Let $t\!=\!0$ be the time of arrival of 
the first photons on the line of sight to the center of the conical 
shell. Photons emitted from a shell's polar angle $\theta$ 
arrive at $t\!=\!R\, (1\!-\!\cos\theta)/c$.  
If the radiation has a power-law spectrum in the shell's rest 
frame, ${\tilde\nu}^{-\beta}$, the spectral 
energy flux seen by the observer has the form (Kumar \& 
Panaitescu~2000)
$F_\nu\! \propto\! \nu^{-\beta}\, \delta^{2+\beta}$,  
where $\delta\!\equiv\!1/\{\gamma\, [1\!-\! (v/c)\, \cos\theta]\}$. Thus, 
for $\gamma\!\gg\! 1$, 
the high latitude emission from a shell collision obeys
$F_\nu\! \propto\! \nu^{-\beta}\, (t+t_0)^{-(2+\beta)}$,
with $t_0= R/(2\, \gamma^2 \, c)$.
Note that the spectral behaviour does not change during 
the temporal power-law decline. 
This is in contradiction with the observed rapid spectral softening.

To confront this problem Liang et al.~(2006) {\it assumed} that the 
high-latitude spectral index $\beta$ is time-dependent but the temporal 
index still satisfies $\alpha(t) \!=\! 2\!+\! \beta(t)$. Although 
structured 
jet models (M\'esz\'aros, Rees \& Wijers~1998; Zhang \& M\'esz\'aros~2002; 
Ross, Lazzati \& Rees.~2002) may yield a time-varying $\beta$, there is 
no reason why it should depend on an angle defined by the position of the 
observer. Indeed, the relation $\alpha(t)\!=\!2\!+\!\beta(t)$ is badly 
violated in canonical light curves (Zhang, Liang \& Zhang 2007).
We conclude that the curvature effect in the currently explored
fireball models does not agree with the data.

\section{The CB model and its predictions}  

In the CB model (e.g., DD2004 and references therein), GRBs and their AGs 
are produced by jets of highly relativistic CBs of ordinary matter (Shaviv 
\& Dar~1995; Dar~1998; Dar \& Plaga~1999). Long-duration GRBs originate 
from CBs ejected in core collapse supernova explosions. The `engine' of 
short GRBs is much less well established, it could be the merger of 
compact objects, e.g.~neutron stars, and/or mass-accretion episodes 
on compact objects in close binaries (e.g., 
microquasars), or even phase transitions of increasingly compactifying 
stars (neutron stars, hyper stars, or quark stars).

The pre-GRB ejecta of the parent stars create `windy' environments of 
`circumburst' material. The early luminosity of the event (a core-collapse 
supernova for long GRBs) permeates this semitransparent material with a 
temporary constituency of scattered, non-radially-directed photons: a {\it 
glory} of visible or UV light, with an approximately `thin-bremsstrahlung' 
spectrum (DD2004). The $\gamma$-rays of a single pulse of a GRB are 
produced as a CB coasts through the glory. The electrons enclosed in the 
CB boost the energy of the glory's photons, via inverse Compton 
scattering, to $\gamma$-ray energies. The initial fast expansion of the 
CBs and the radially-increasing transparency of the windy environment 
result in the exponential rise of a GRB pulse.  As a CB proceeds, the distribution of 
the glory's light becomes more radially directed, its density decreases. 
Consequently, the energy of the observed photons is continuously shifted 
to lower energies as their number plummets. These trends were observed in 
CGRO/BATSE data (Giblin et al.~2002; Connaughton~2002; Ryde \& 
Svensson~2002, DD2004).  During a GRB pulse the spectrum softens and the 
peak energy decays with time as a power law. This is also the behaviour of 
the X-ray flares of a GRB, which are either the low-energy tails of 
$\gamma$-ray pulses, or fainter and softer signals with the same origin 
(DDD2007a). Typically, the fast decline of the prompt emission in the 
$\gamma$-ray and X-ray bands is taken over, within few minutes of 
observer's time, by the `afterglow' --synchrotron emission from swept-in 
ISM electrons spiraling in the CB's enclosed magnetic field.

The above effects can be explicitly analized (DD2004), and 
summarized to a good approximation
in a {\it master formula} (DDD2007a)
 for the temporal shape and spectral evolution of the energy fluence
of an ICS-generated $\gamma$-ray pulse (or X-ray flare):
\begin{equation}
F_E^i\propto E\, {d^2N_\gamma^i\over dt\,dE} \propto
\Theta[t-t_i]\;
e^{-[\Delta t_i/(t-t_i)]^m}\,
 \left\{1-e^{-[\Delta t_i/(t-t_i)]^n}\right\}\, E\,{dN_\gamma^i(E,t) \over dE}\; ,
\label{GRBlc}
\end{equation}
where `{\it i}' denotes the i-th pulse, produced by
a CB launched at (an observer's) time $t_i$. In Eq.~(\ref{GRBlc}),
the time scale is set by $\Delta t_i$, with $\gamma\, \delta\, c\, \Delta t_i/(1+z)$  
the radius of transparency of the glory, within which its photons are
approximately isotropic. In $\Delta t_i$ time units, a pulse rises 
as $Exp[-1/t^m]$, $m\!\sim\!1$ to 2, and decreases as $1/t^n$, $n\!\sim\!2$. Finally,
$E\, dN_\gamma^i/dE$ is the spectral function of the glory's 
photons, up-scattered by  the CB's electrons, and discussed anon.

The glory has a thin thermal-bremsstrahlung spectrum:
$\epsilon\, {dn_\gamma / d\epsilon} \!\sim\!
(\epsilon/\epsilon_g)^{1-\alpha_g}\,e^{-\epsilon/\epsilon_g}$,
with a typical (pseudo)-temperature
$ \epsilon_g \!\sim\!1$ eV, and index $\alpha_g\!\sim\!1$.
During the $\gamma$-ray phase of a GRB,
the Lorentz factor $\gamma$ of a CB stays put at its initial value,
for the deceleration induced by
the collisions with the ISM has not yet had a significant effect
(DDD2002, DD2007).
Let $\theta$ be the observer's angle relative to the direction of motion
of a CB and let the corresponding Doppler factor be
$\delta\!=\!1/ \{\gamma\,[1\!-\!\,(v/c)\, \cos\theta]\}$. 
Let $\theta_i$ be the angle of incidence of the initial
photon onto the CB, in the parent star's rest system.
The energy of an observed photon, 
Compton scattered in the glory by an electron comoving with a
CB at redshift $z$, is given by
$E\!=\!\gamma\, \delta\, \epsilon \, (1\!+\!\cos\theta_i)/(1+z)$.
The predicted GRB prompt spectrum is (DD2004):
\begin{equation}
E\, {dN\over dE} \sim \left({E\over T}\right)^{1-\alpha_g}\,
 e^{-E/T}+ b\,(1-e^{-E/T})\, \left({E \over T}\right)^{-p/2} .
\label{GRBspec}
\end{equation}
The first term, with $\alpha_g\!\sim\! 1$, is the result of Compton
scattering by the bulk of the CB's electrons, which are comoving with it.
The second term in Eq.~(\ref{GRBspec}) is induced by
a very small fraction of
`knocked on' and Fermi-accelerated electrons, whose initial spectrum
(before Compton and synchrotron cooling) is $dN/dE_e\propto E_e^{-p}$,
with $p\approx 2.2$. Finally, $T$ is the effective (pseudo)-temperature
of the GRB's photons:
\begin{equation}
T\equiv{4\, \gamma\, \delta\,\epsilon_g\,
\langle 1+\cos\theta_i\rangle / [3\, (1+z)]}\,.
\label{ICST}
\end{equation}
For a semi-transparent glory $\langle\cos\theta_i\rangle$ would be
somewhat smaller than zero.

For $b={\cal{O}}(1)$,
the energy spectrum predicted by the CB model, Eq.~(\ref{GRBspec}),
bears a striking resemblance
to the Band function (Band et al.~1993) traditionally used to model the
energy spectra of GRBs. For many Swift GRBs the spectral observations
do not extend to energies much bigger than $T$, or the value of $b$
in Eq.~(\ref{GRBspec}) is relatively small, so that the first term
of the equation provides a very good approximation.
This term coincides with the `cut-off
power-law' spectrum recently used to model
GRB spectra. It yields a `peak-energy' (the maximum of $E^2\, dN/dE$
at the beginning of a pulse)
$E_p\!=\!(2\!-\!\alpha_g)\, T\!\approx\! T$ for
$\alpha_g\!\sim \!1$. At later times, the CB is sampling the glory at distances
for which its light is becoming increasingly radial, 
$\langle1\!+\!\cos\theta_i\rangle\!\to\!1/r^2\!\propto\!1/t^2$ in Eq.~(\ref{ICST}).
The value of $E_p(t)$ consequently decreases as: 
\begin{equation}
E_p(t)\approx E_p(t_i)\,
\left[1-{t-t_i\over \sqrt{\Delta t_i^2+(t-t_i)^2}}\right]\, .
\label{Epi}
\end{equation}
The light-curve generated by a  sum of pulses is well 
approximated (DDD2007a) by:
\begin{equation}
F_E
\approx \sum_i\,A_i\, \Theta[t-t_i]\;
e^{-[\Delta t_i/(t-t_i)]^2}\,
 \left\{1-e^{-[\Delta t_i/(t-t_i)]^2}\right\}\,\left[{E/ 
E_p(t)}\right]^{1-\alpha_g} \
e^{-E/E_p(t)}\,
\label{GRBXlc}
\end{equation}
until ICS is overtaken by synchrotron radiation.

In   X-rays the distinction between a prompt and an afterglow period
can be made precise, they correspond to the successive
dominance of the two radiation mechanisms: ICS and SR.
The  actual form of the SR-dominated
AG spectral energy flux, $F_\nu$,
we have discussed very often (DDD2007a,b and references therein).  
Suffice it to recall that (for cases as the ones we discuss here,
whose AG can be well fit with a single dominant or average CB)
the shape of the observed $F_\nu$, corrected for absorption, is 
determined by 
$\gamma_0\,\theta$, and its time scale is
determined by a deceleration time, $t_0$, at which $F_\nu$
achromatically `bends down' towards its asymptotic behaviour,
$F_\nu \propto \nu^{-\beta}\, t^{-\beta-1/2}$.
Typically $\beta\sim 1.1$ (DDD2007b).
 
\subsection{The hardness ratio in the CB model}

For a case in which the X-ray-absorption  factor $A(E)$ is known,
we have given enough information to predict the hardness ratio (HR)
from the X-ray energy flux  in a given energy band. For the late
SR-dominated phase, this is trivial. A look at an X-ray light curve, such that of
Fig.~\ref{f1}a, tells one the time at which the fast-decline ends, meaning
that SR starts to dominate. From that time onwards, the HR
is that corresponding to  $F_E \!\propto\! E^{-\beta}$, which would be 
HR $\approx\! 0.18$ for an
unabsorbed flux with $\beta\!=\!1.1$.

In the ICS-dominated phase, the shape of the flux determines 
the number of flares one ought to fit, one in Fig.~\ref{f1}a, for instance. 
If one uses Eq.~(\ref{GRBXlc}) with $\alpha_g\!=\!1$, each pulse 
is fit with 4 parameters:  $t_i$,  $\Delta t_i$, 
$E_p(t_i)$, and $A_i$. For the rapid-decline phase, it
suffices to consider the main or the latest few flares, since 
the last factor in Eq.~(\ref{GRBXlc}) suppresses 
 the relative contribution of earlier flares by the time 
the data sample the later ones. Once the parameters are
fixed, the HR is determined by the quotient of the integrals
$\int dE \,A(E)\,d^2N_\gamma/dEdt$ in the two Swift X-ray energy bands.
This rosy picture is clouded by two facts: the data for the {\it integrated}
flux in the 0.3-10 keV interval is very insensitive to the values of 
$E_p(t_i)$, which the fit consequently returns with very large errors;
we do not know $A(E)$.

We studied numerically the HR of a pulse given
by Eq.~(\ref{GRBXlc}), in the large interval $0\!<\!t\!-\!t_i\!<\!10\,\Delta t_i$,
for an exaggerated  range of $E_a$ in $A(E)\!\approx\!Exp[-(E_a/E)^3]$.
We found that
\begin{equation}
{\rm HR}_i(t)=B\,e^{-\;\left[{\Delta E/ E_p(t_i)}\right]\,
\left[1-(t-t_i)/\sqrt{\Delta t_i^2+(t-t_i)^2}\,\right]^{-1}}\; ,
\label{HR}
\end{equation}
is a fair approximation, with $\Delta E$ an effective interval between the bands in the HR. 
More explicitly, if $B$ and $\Delta E/E_p(t_i)$ are {\it fit}, the approximation
is good to a few \% for a typical $E_p(t_i)\!>\! 200$ keV, deteriorating
to $\sim\!40$\% for an extreme and atypical $E_p(t_i)\!=\!30$ keV.
We shall consequently fit $B$ and $\Delta E/E_p(t_i)$ in comparing theory and data for the HR.

For times at which the late-time tail of a single pulse dominates, the HR satisfies
\begin{equation}
{\rm HR}_i(t) \to B\,e^{-\left[\Delta E / E_p(t_i)\right]\,
\left[2\,(t-t_i)^2/ \Delta t_i^2\right] }
\label{HRapprox}
\end{equation}
with  precision increasing with $t$.

\section{Hardness ratios: case studies}

The Burst Alert Telescope (BAT) on Swift has detected nearly 250 GRBs or 
XRFs whose X-ray emission was followed with its X-Ray Telescope (XRT) from 
$\sim\!70$ s after trigger until it faded away. Incapable of discussing 
all these observations, we first study five cases, which we view as representative, 
and which have well-sampled X-ray fluxes and hardness ratios during the 
fast-decline and the ensuing AG phase. They are: the `clean' single-peak 
XRF 060218, GRB 060904a with its 4 X-ray flares during the fast decline 
phase, the simpler two-flare GRB 061121, the duller GRB 061126, for 
which the XRT observations began late and the bright GRB 061007
with an approximate single power-law afterglow.

\noindent 
{\bf XRF 060218:} 
This single-peak XRF provides one of the best testing grounds of theories, 
given its proximity, which resulted in very good sampling and statistics. 
The BAT data lasted 300s, beginning 159s after trigger, with most of the 
emission below 50 keV (Campana et al.~2006; Liang et al.~2006). The 
prompt X-ray emission lasted more than 2000s, during which the peak energy 
evolved from 54 keV to $\leq 5$ keV at the start of the fast-decline 
phase. The flux and HR data are shown in Figs.~\ref{f1}a,c. During the afterglow 
phase, the HR seems to decrease gradualy from $\sim\! 0.8$ at 6.2 ks 
to $\sim\!0.25$ at 72 ks. In the CB model such a trend could be 
produced by diminishing absorption along the line of sight to the CBs. 

The HR 
from unabsorbed synchrotron radiation with a typical $\alpha\!=\!1.1$ is
HR $\approx 0.18$, well below the reported HR for the absorbed flux 
of XRF 060218 (Evans et al.~2007).  
Pian et al.~(2008) reported that the extinction derived 
from the equivalent width of the Na I D absorption line in the spectrum of 
the associated SN2006aj is $E(B-V)\!=\!0.13\pm 0.02$, consistent with 
Galactic extinction and no 
extinction in the host. The H column density needed to fit 
the Swift X-ray prompt spectrum was NH = 
$6\times 10^{21}\, {\rm cm^{-2}}$ (Campana et al.~2006), implying 
$E(B-V)\!\approx\! 1$, considerably greater than the total 
extinction (in the Galaxy, the intergalactic medium and the host galaxy)
derived from optical emission and absorption lines, as 
well as from the optical colours of the afterglow, measured by Mirabal et 
al.~(2006). These authors stress that NH is not really NH, but a 
proxy for the heavier elements that dominate the X-ray photoelectric 
absorption, and that the relatively small extinction implies a 
dust-deficient medium such as the stellar wind of a Wolf-Rayet progenitor, 
which has enough column density to be the location of this excess 
photoelectric X-ray absorption and relatively dust-deficient medium. The 
use of this large NH deduced from the prompt emission to infer the 
late X-ray afterglow spectrum may have resulted in the very large 
$\Gamma\!\approx\! 4.4\pm 1.0$ reported in 
{\it http://swift.physics.unlv.edu/xrt.html}.

\noindent
{\bf GRB 060904a:} 
The BAT detected a weak emission of $\gamma$ rays for about a 
minute, with several small peaks before the main burst,
also seen by the Konus-Wind and 
Suzaku satellites (Yonetoku et al.~2007). The XRT  
followed the fast decline of the main burst and saw three additional 
flares, as shown in Fig.~\ref{f1}b. 
A rapid spectral softening was observed during both the prompt tail
phase and the decline phase of the X-ray flares, see Fig.~\ref{f1}d.
Due to a second GRB (0060904b) being 
detected just 1.5 hours later, Swift slewed away from GRB 060904a, so that
there were no data during a couple of hours until the 
XRT returned  to follow its fading afterglow.
After correcting for absorption (Yonetoku et al.~2007), 
the photon spectral index during the AG phase was found to be 
$\Gamma\! =\!2.1\pm 0.1$.

\noindent
{\bf GRB 061121:}
The $\gamma$-ray burst started with a bright precursor which 
lasted 10s. Then, 50s later, there was a much brighter 
burst of $\gamma$ rays. Swift had already turned its XRT when the 
second $\gamma$-ray flare occurred and  the X-ray emission was measured
during the actual event and its subsequent rapid decline, as shown in
Fig.~\ref{f2}a. After the rapid decline, the photon spectral index, corrected for
absorption, was $\Gamma\!=\!2.05\pm 0.15$ (Page et al.~2007).

\noindent {\bf GRB 061126:} This very long burst had four main overlapping 
peaks, the last peak ending $\sim\!25$s after trigger, but low-level 
emission was detected until $\sim\!200$s later. The RHESSI satellite also 
detected this burst, and also saw $\gamma$-ray emission for $\sim\!25$s. 
The XRT detected the X-ray emission only long after the prompt emission 
had faded. These late data are shown in Fig.~\ref{f2}b. The photon spectral index 
after correcting for absorption (Perley et al.~2007) is $\Gamma=2.00\pm 
0.07$, and is time-independent, suggesting that the entire XRT light curve 
is that of the synchrotron afterglow of GRB 061126.

\noindent {\bf GRB 061007:} 
This long bright burst lasted $75\pm 5$ s. Its lightcurve showed three 
large peaks, and a smaller peak starting at 75 s, rising to a maximum 
at 79 s and declining with a very long and fast 
decay. The XRT began follow-up observations 80 s after trigger. The 0.3--10 
keV light curve (Fig.~\ref{f3}a) shows a single power-law decline with a slope of 
$1.6\pm 0.1$. In the CB model this is the tail of a 
cannonical AG whose `plateau' ended before the XRT began its 
observations. The predicted photon spectral index (DDD2007b), 
$\Gamma=\alpha+0.5=2.10\pm 0.10$, is consistent with the best fit 
spectral index, $\Gamma=2.03\pm 0.10$, shown in Fig.~\ref{f3}d.

\subsection{Hardness ratios: CB-model results}
In the CB-model the SR-dominated X-ray afterglow, if corrected for 
absorption, has a time-independent photon spectral index, 
$\Gamma\!\sim\!2.1$, and 
a constant hardness ratio. This expectation is consistent, within 
observational errors, with the Swift data in all the cases we considered,
with the possible exception of XRF 060218, whose complex situation
regarding absorption corrections we have reviewed.
The spectral behaviour is much more complex during the prompt emission.

Since XRF 060218 is a single-flare event, its light curve and the 
evolution of its HR, shown in Figs.~\ref{f1}a,c, are simple. The agreement 
between the model expectations and the XRT observations is satisfactory. 
The CB-model parameters are specified in Table~1. Multi-flare events such 
as GRB 060904a and, to a lesser extent, GRB 061121, require 
multi-parameter fits; the number of peaks we fit and their relevant 
parameters are also specified in Table~1. The way the HR of these 
bursts 
predictably follows the ups and downs of the flux is quite impressive, 
compare Fig.~\ref{f1}b with \ref{f1}d, and 
Fig.~\ref{f2}a with \ref{f2}c. For GRBs 061126 (figs.~\ref{f2}b,c and \ref{f3}b)
and 061007 (figs.~\ref{f3}a,c,d), 
the available data covers only the SR-dominated X-ray AG where, as 
expected, the HR ratio is constant. Note in Fig.~\ref{f2}b that, although the 
late time behaviour of the flux has the shape predicted by the CB model, 
the measured points lie systematically above the prediction. Such a 
discrepancy may result from a decreasing X-ray absorption along the line 
of sight to the AG source. The fluxes reported in the SWIFT XRT repository 
(Evans et al. 2007) assume a constant absorption during the entire 
measurements.  In the CB model, the jet of CBs moves hundreds of parsecs 
during the observations, and the absorption may decrease with time as the 
jet approaches the halo of the host galaxy.

\section{CB-model results for the effective spectral index}

The spectral index, $\Gamma(t)\!=\!\beta(t)+1$ of many GRBs,
extracted from an empirical power-law parametrization,
$ F_\nu\!\propto\! \nu^{-\beta}\,t^{-\alpha}$, is reported in
{\it http://swift.physics.unlv.edu/xrt.html},
and discussed in more detail for a selected set of bright GRBs by
Zhang et al.~(2007). As reported in the
introduction, these results on $\Gamma(t)$ may themselves be a rough description 
of rapidly-varying spectra potentially having an exponential energy-dependence,
as in Eq.~(\ref{GRBXlc}). Yet, we may define an {\it effective} index via the logarithmic  
derivative of the prompt ICS spectrum. For a single pulse in Eq.~(\ref{GRBXlc}):
\begin{equation} 
\Gamma_{\rm eff}(E,t-t_i) =-E\; {d\,{\rm log} F_E\over dE}\Bigg|_{E=\widetilde E}=
                  \alpha_g+{\widetilde E\over E_p(t-t_i)}\,,
\label{Geff}
\end{equation}
where ${\widetilde E}$ is an effective constant energy, $t$ is the time
after trigger, and $\alpha_g\!\approx\! 1$ is defined in Eq.~(\ref{GRBspec}).
For the synchrotron afterglow, the CB model predicts a power-law 
spectrum with roughly a constant photon index $\Gamma_{\rm SR}$, 
and a late-time temporal power-law decline with a power-law index 
(DDD2007b):
\begin{equation}
\alpha= \Gamma_{\rm SR}-1/2.
\label{alpha}
\end{equation}

In the data analysis in
{\it http://swift.physics.unlv.edu/xrt.html}, for lack of sufficiently
large statistics, different time intervals were coadded, smoothing the time-dependence 
of the fitted spectral index.
For an `effective-index' study of the results of this data analysis, a
single-pulse approximation is adequate to the description of a GRB's 
$\Gamma(t)$ at the end of the prompt phase and during the fast decline.
In this approximation, for a pulse starting at $t\!=\!t_i$, followed by a 
SR-dominated afterglow, the rough CB-model prediction is:
\begin{equation}
\Gamma_{\rm eff}\sim \left[1+ 
{\widetilde E \over E_p(t)}\right]\, \Theta[t_{\rm AG}-t]\,\Theta[t-t_i]+ 
\Gamma_{\rm SR}\, \Theta[t-t_{\rm AG}]\, ,
\label{Gammat}
\end{equation}
where  $t_{\rm AG}$ is the time at 
which the SR `afterglow' takes over the ICS `prompt' emission.
The assumed rather abrupt transition from the ICS-dominated
first term in Eq.~(\ref{Gammat}), to the second SR-dominated term,
is justified by Eqs.~(\ref{Epi}, \ref{GRBXlc}). Indeed, the late decline
of the ICS-dominated term is exponential in the square of the time.

\begin{figure}[]
\centering
\vbox{
\hbox{
\epsfig{file=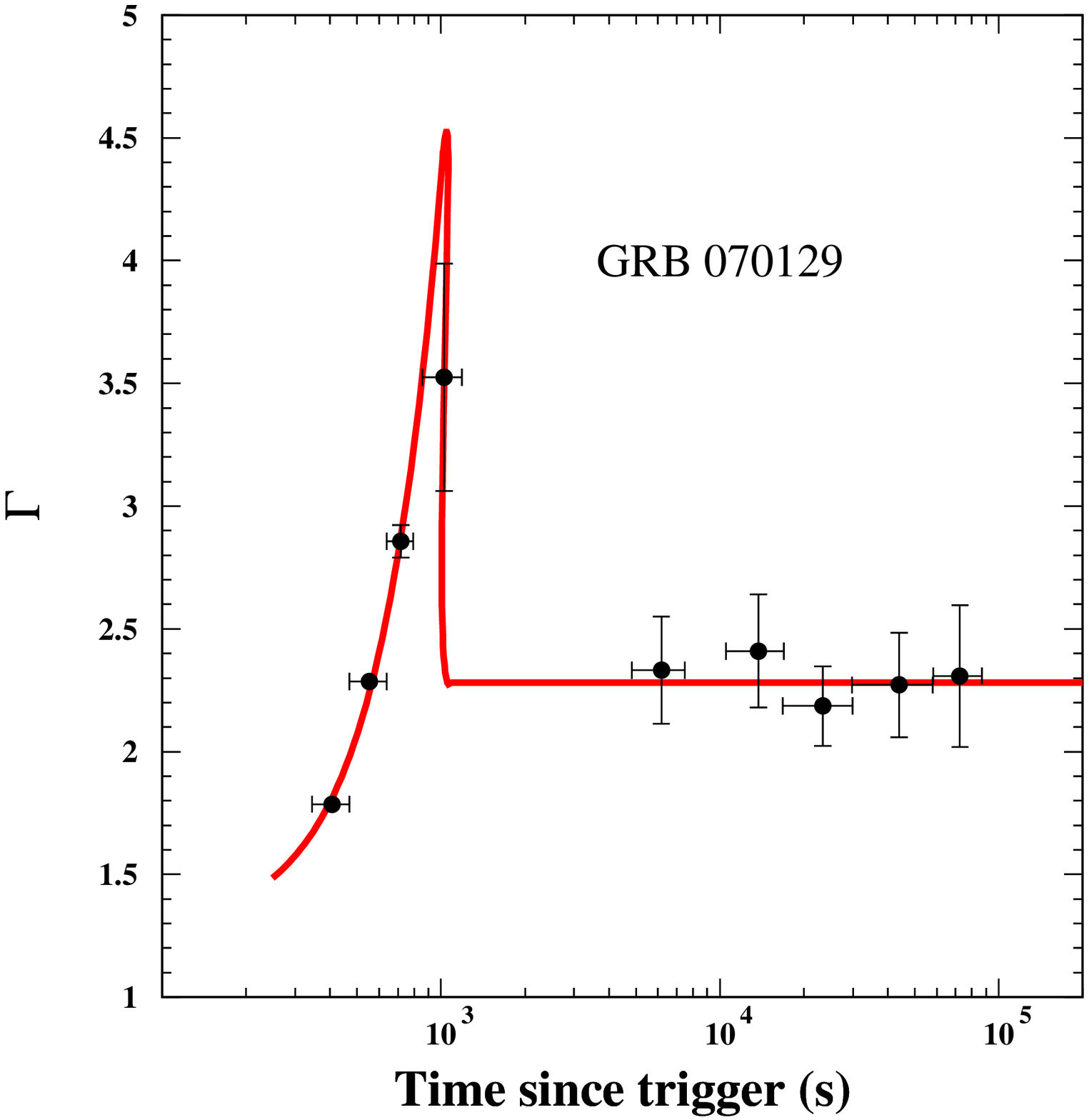,width=8.0cm}
\epsfig{file=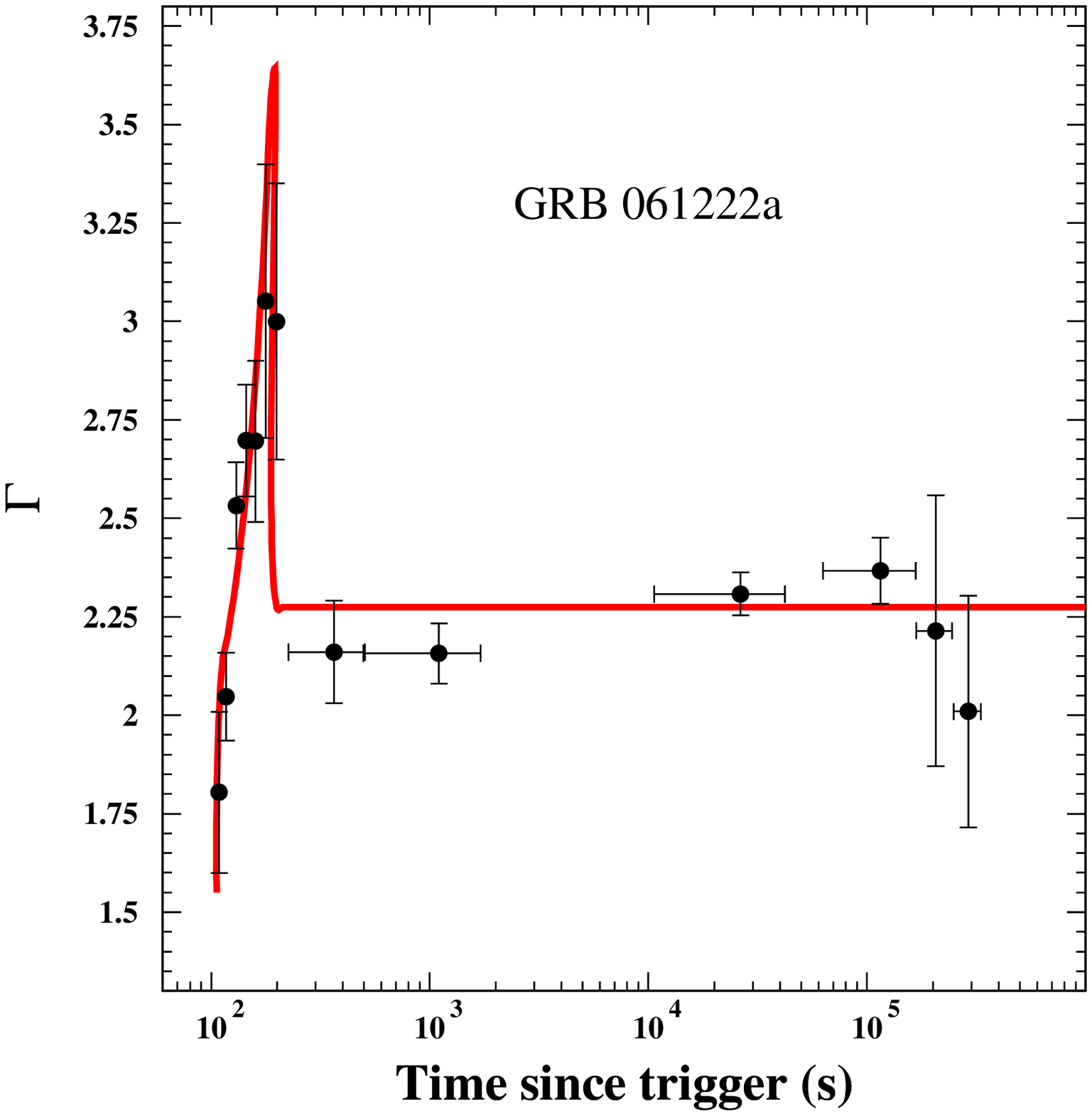,width=8.0cm}
}}

\vbox{
\hbox{
 \epsfig{file=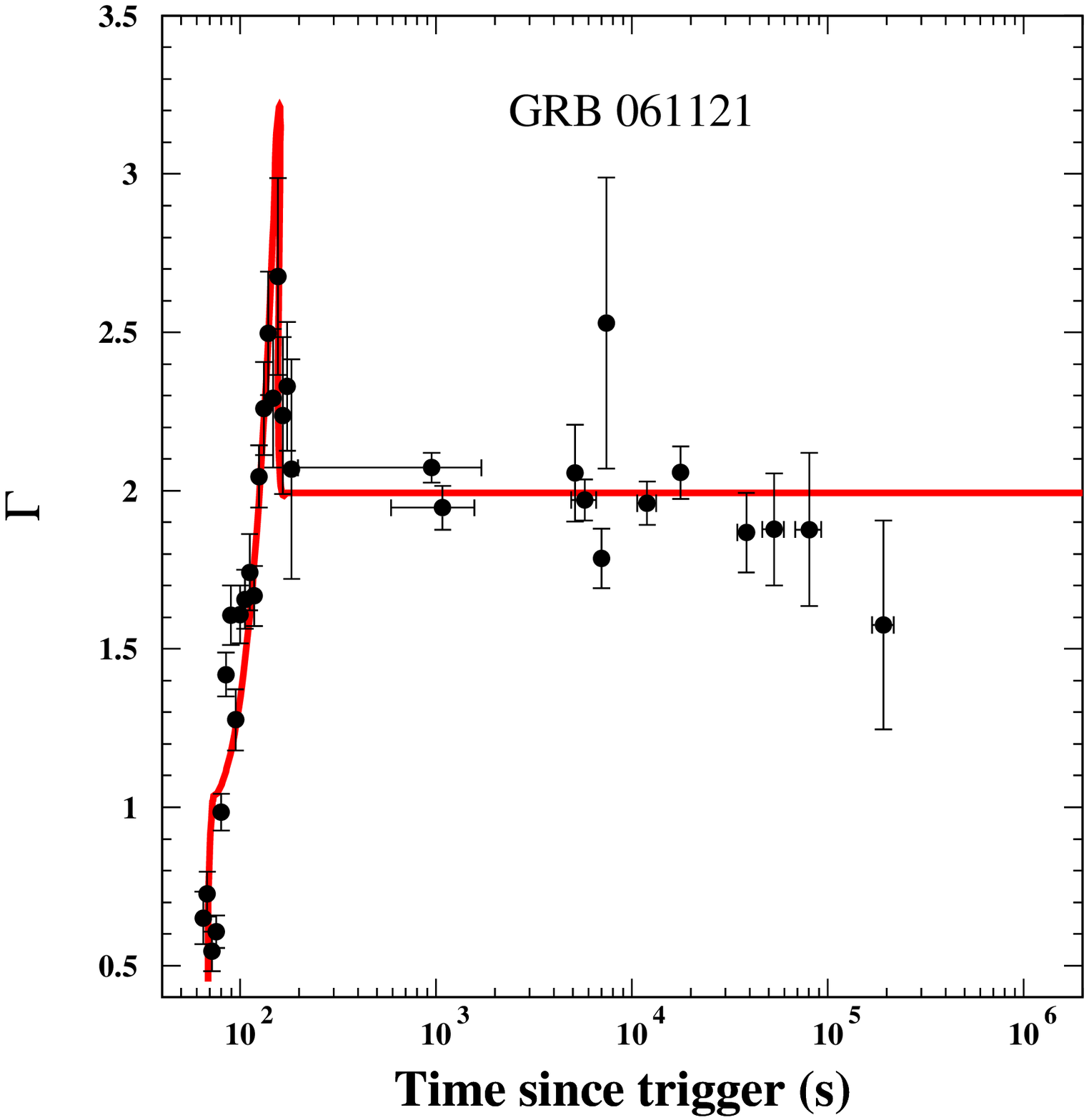,width=8cm}
 \epsfig{file=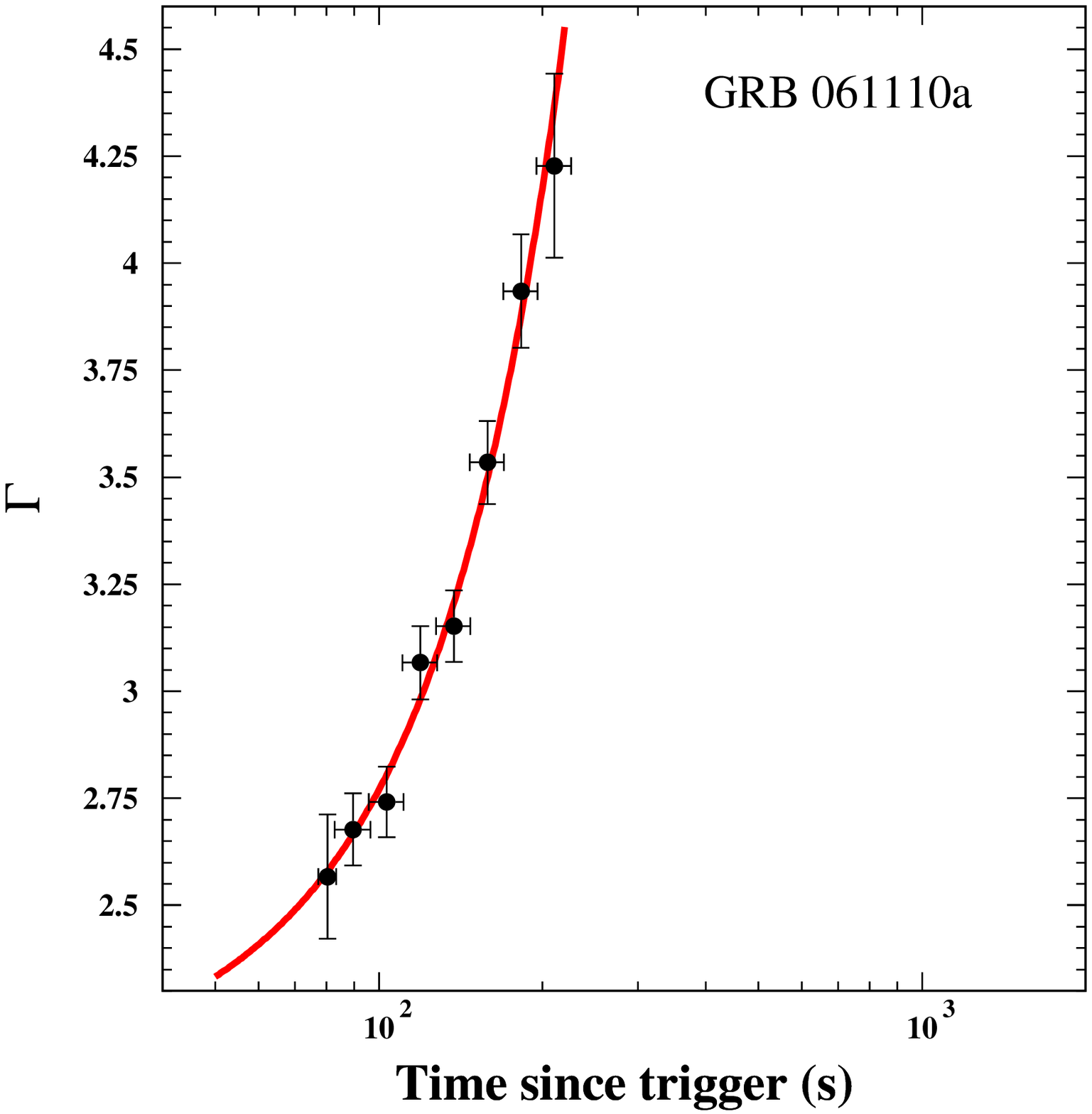,width=8cm}
}}
\caption{
Comparisons between the effective photon spectral 
index in the 0.3-10 keV X-ray band as inferred from 
observations of GRBs with the Swift XRT, 
and the CB model approximate prediction, Eq.~(\ref{Gammat}).
{\bf Top left (a):} GRB 070129.
{\bf Top right (b):} GRB 06122A.
{\bf Bottom left (c):} GRB 061121.
{\bf Bottom right (d):} GRB 061110A.
$\Gamma$ values are from {\it http://swift.physics.unlv.edu/xrt.html}.
}
\label{f4}
\end{figure}

\begin{figure}[]
\centering
\vbox{
\hbox{
\epsfig{file=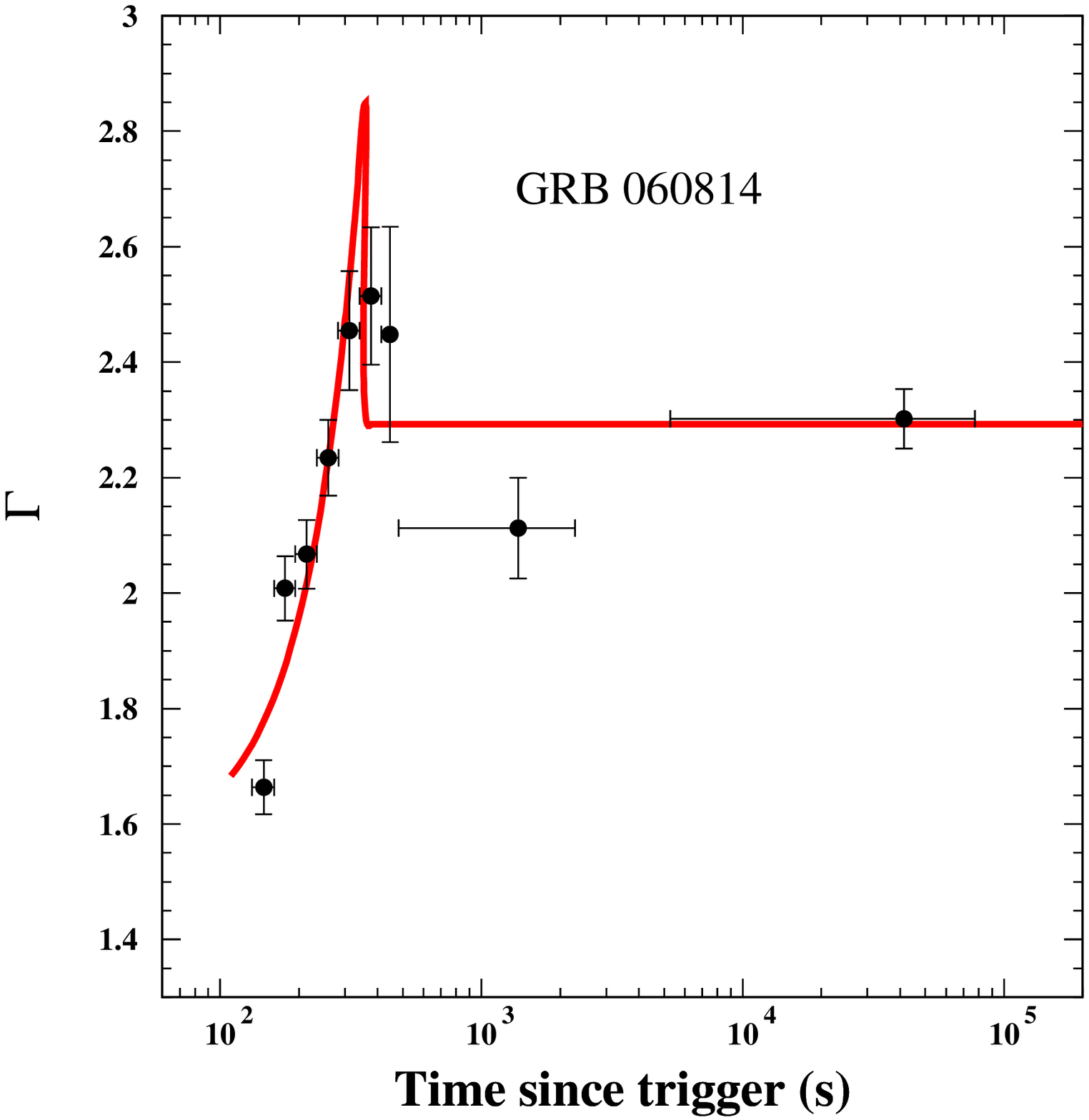,width=8.0cm}
\epsfig{file=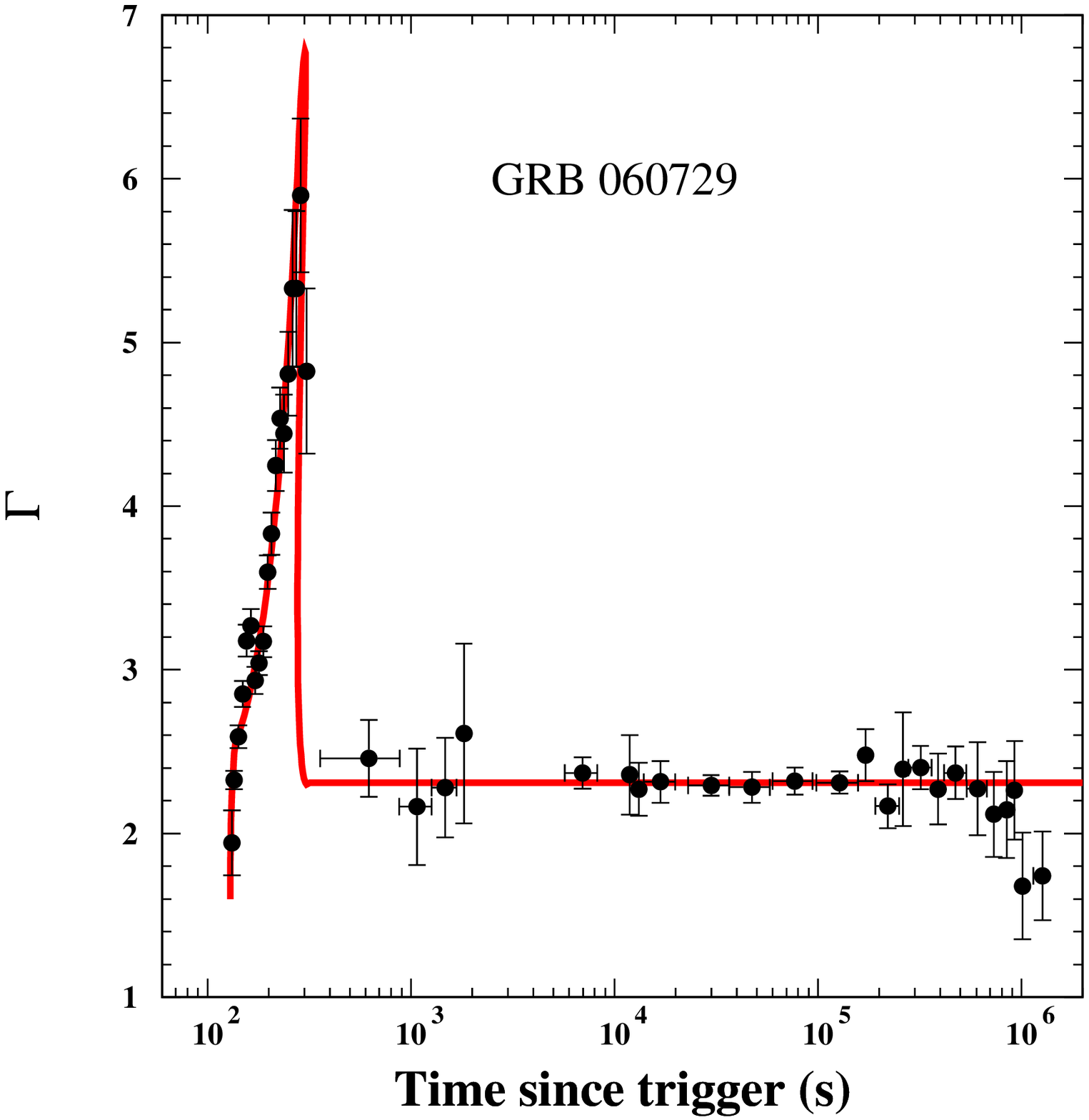,width=8.0cm}
}}

\vbox{
\hbox{
 \epsfig{file=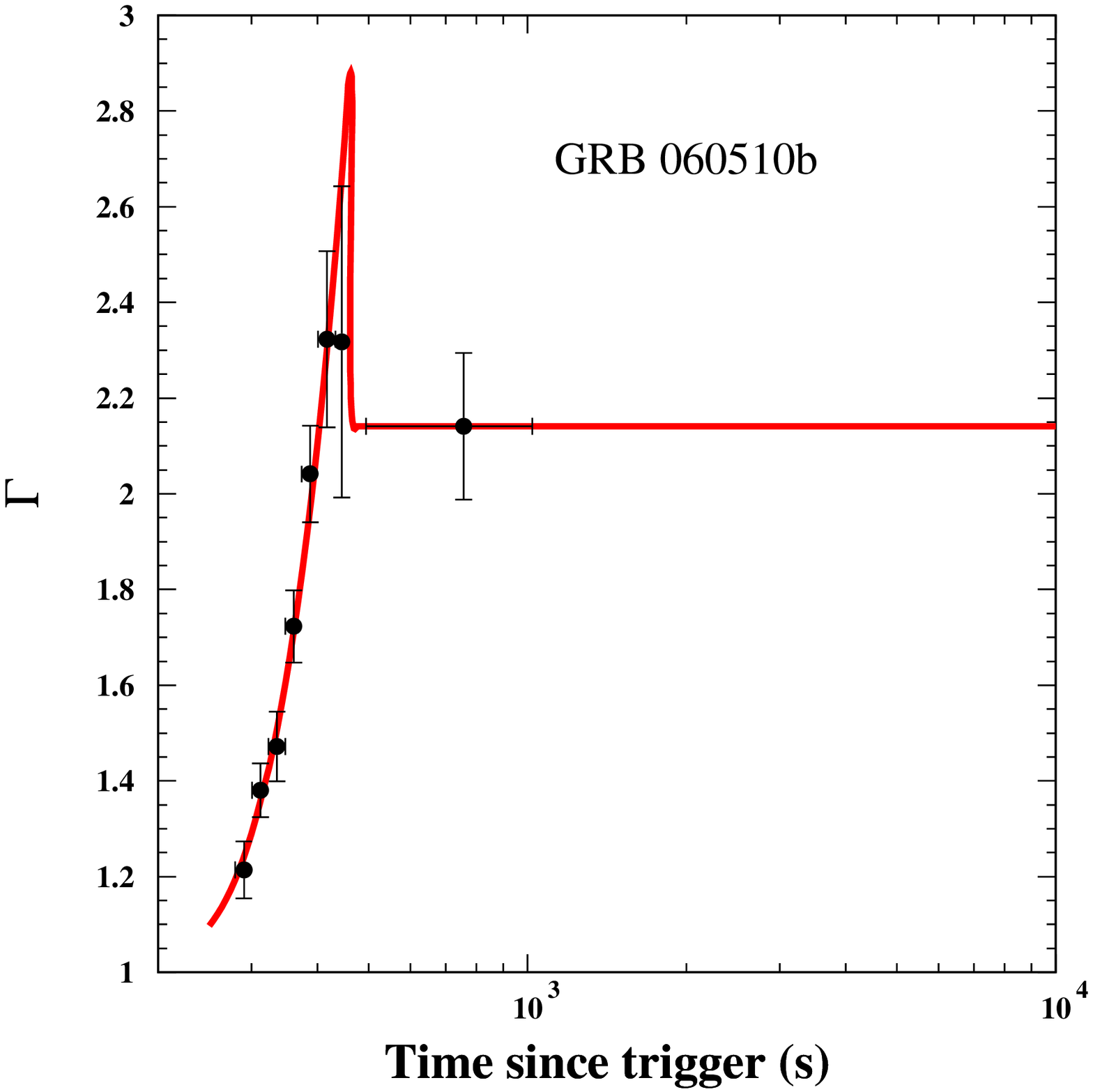,width=8cm}
 \epsfig{file=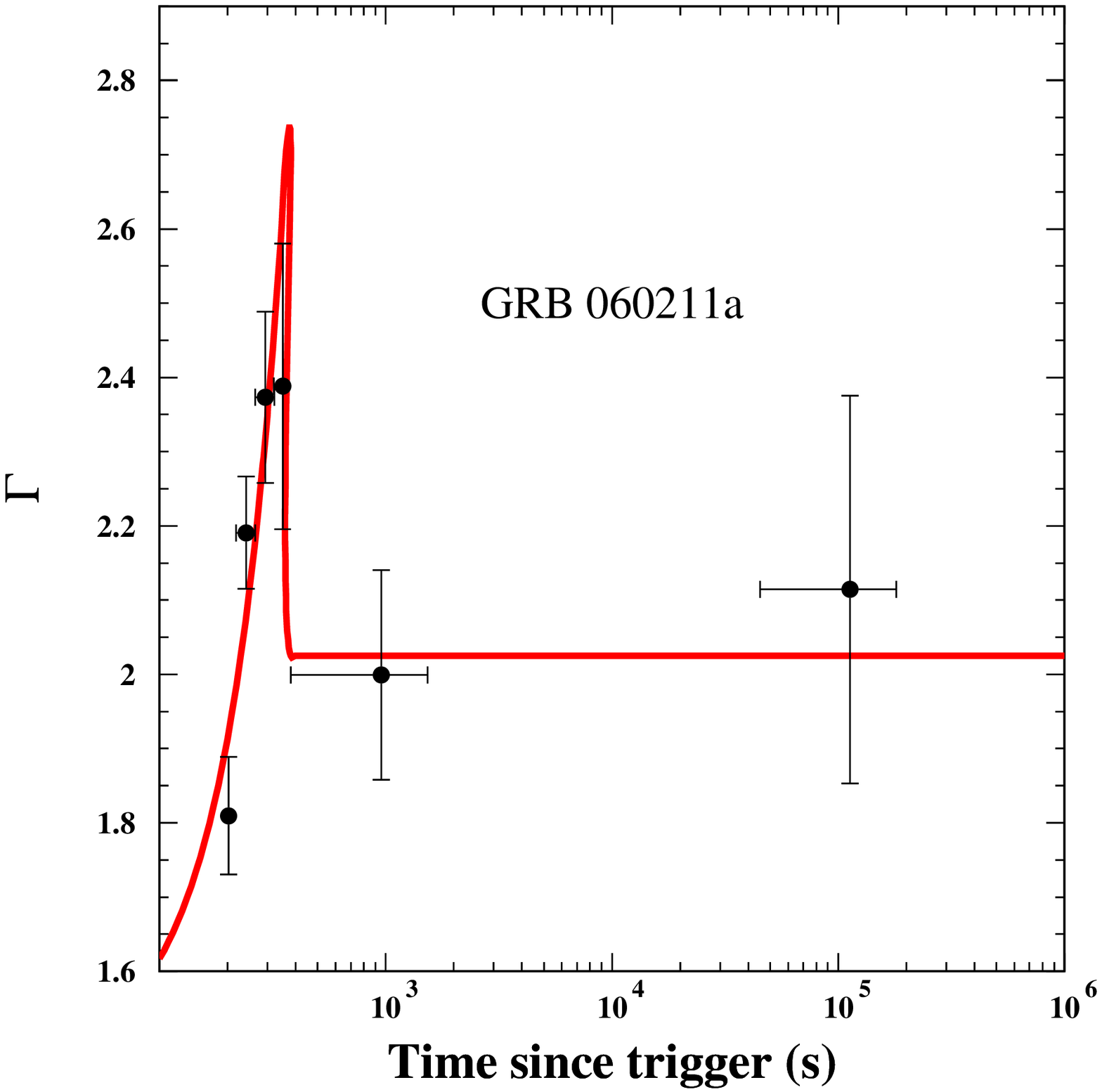,width=8cm}
}}
\caption{
Comparisons between the effective photon spectral
index in the 0.3-10 keV X-ray band as inferred from
observations of GRBs with the Swift XRT,
and the CB model approximate prediction, Eq.~(\ref{Gammat}).
{\bf Top left (a):} GRB 060814.
{\bf Top right (b):} GRB 060729.
{\bf Bottom left (c):} GRB 060501B.
{\bf Bottom right (d):} GRB 060211A.
$\Gamma$ values are from {\it http://swift.physics.unlv.edu/xrt.html}.
}
\label{f5}
\end{figure}

\begin{figure}[]
\centering
\vbox{
\hbox{
\epsfig{file=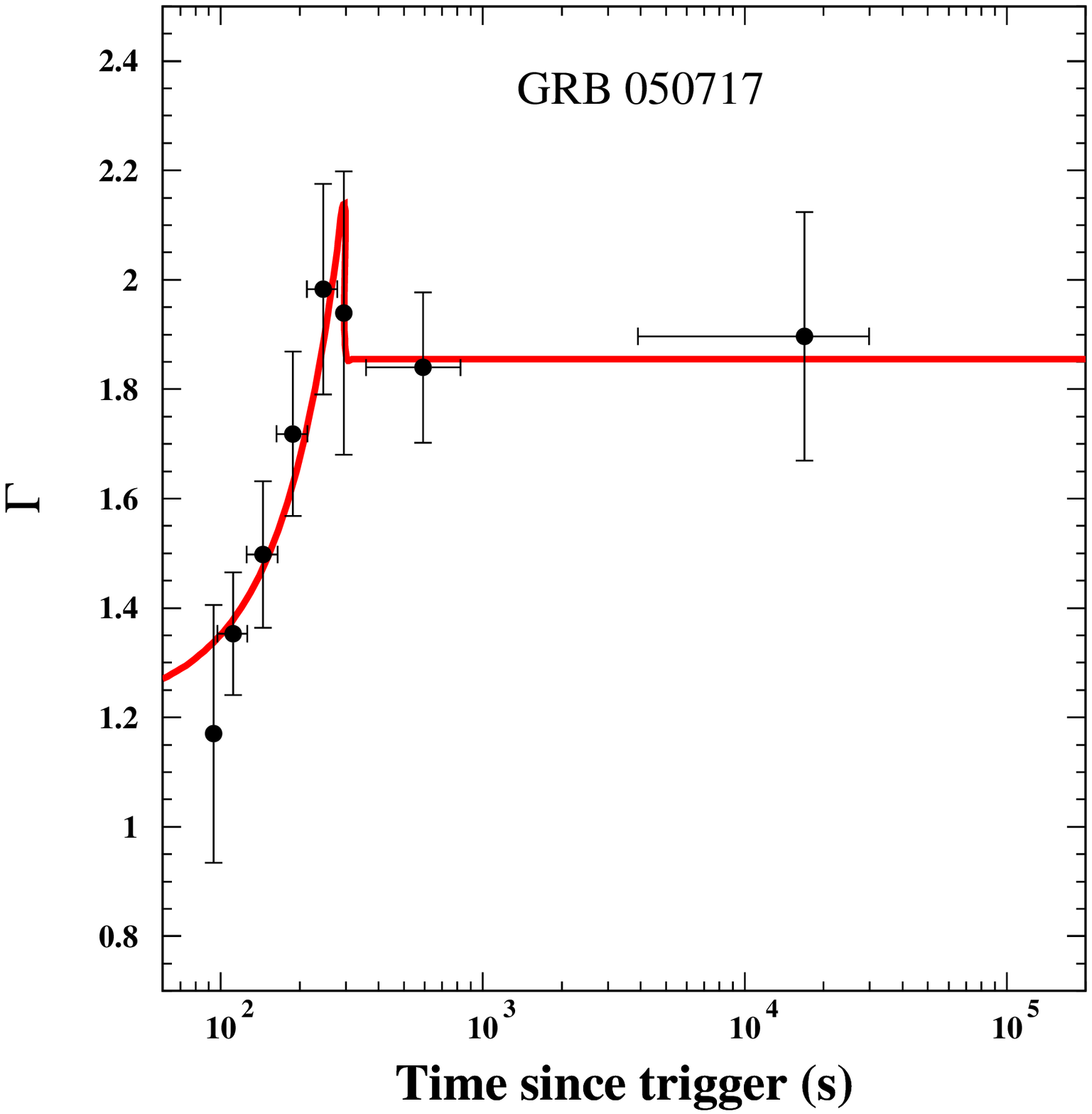,width=8.0cm}
\epsfig{file=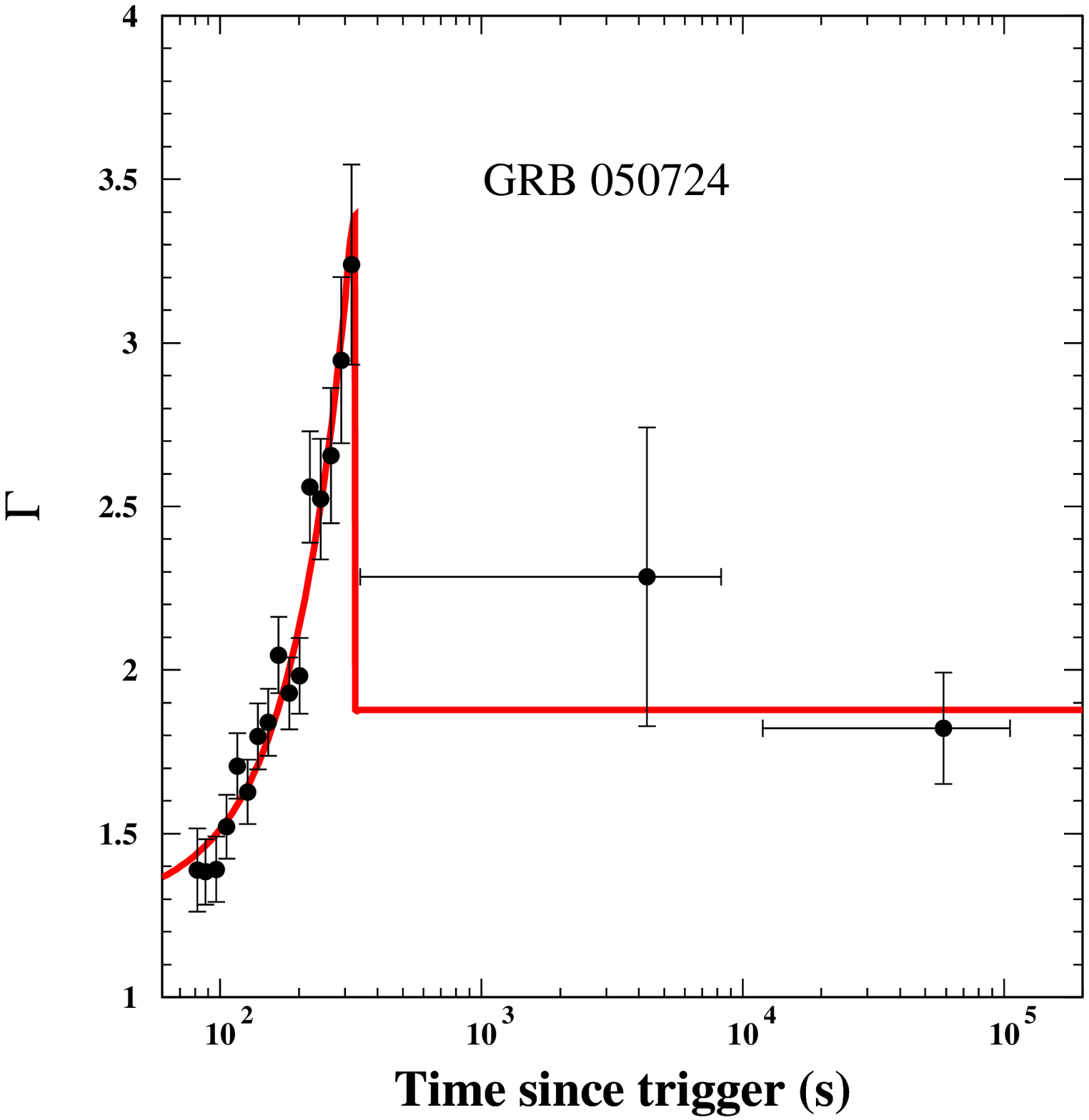,width=8.0cm}
}}

\vbox{
\hbox{
 \epsfig{file=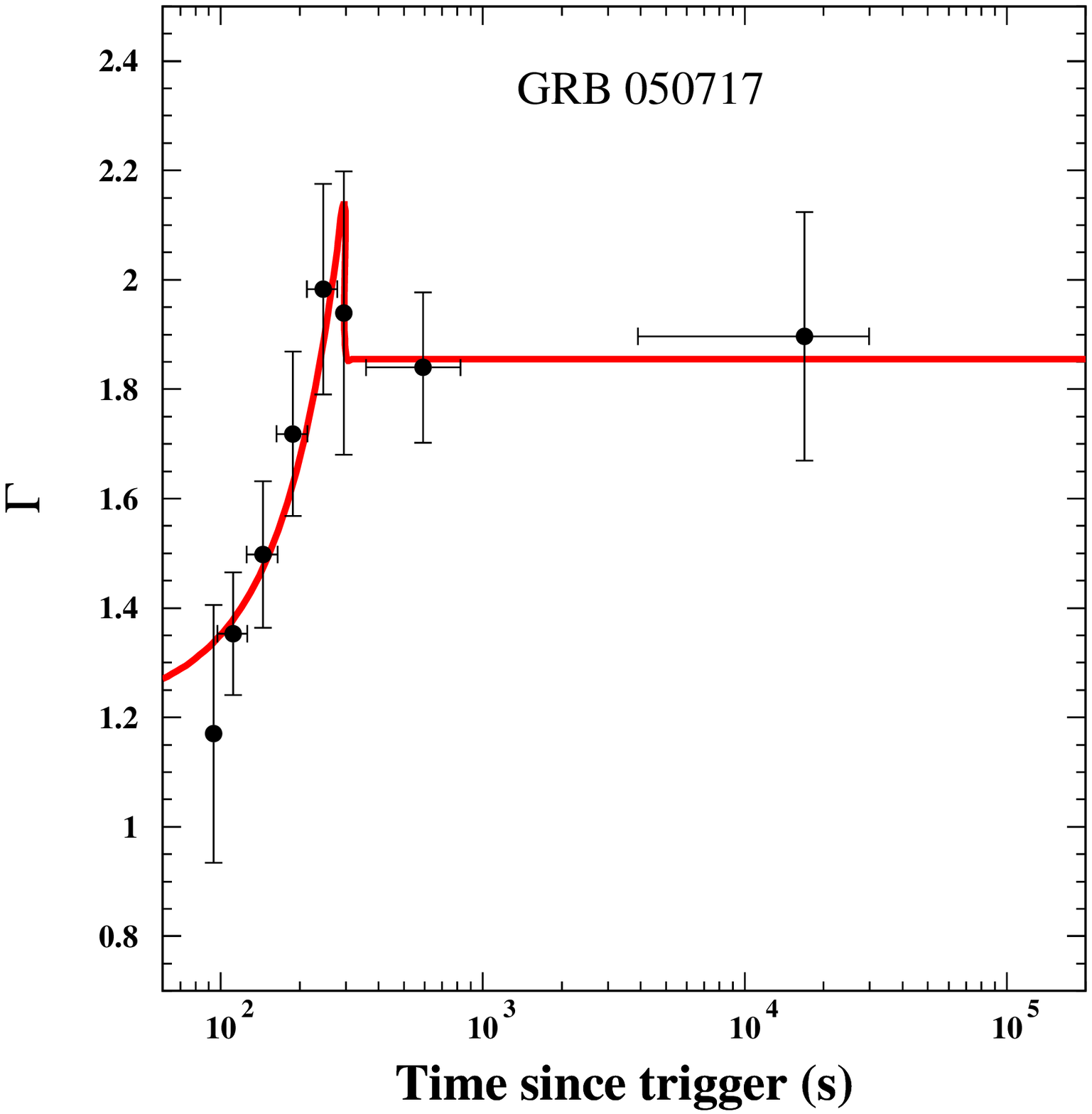,width=8cm}
 \epsfig{file=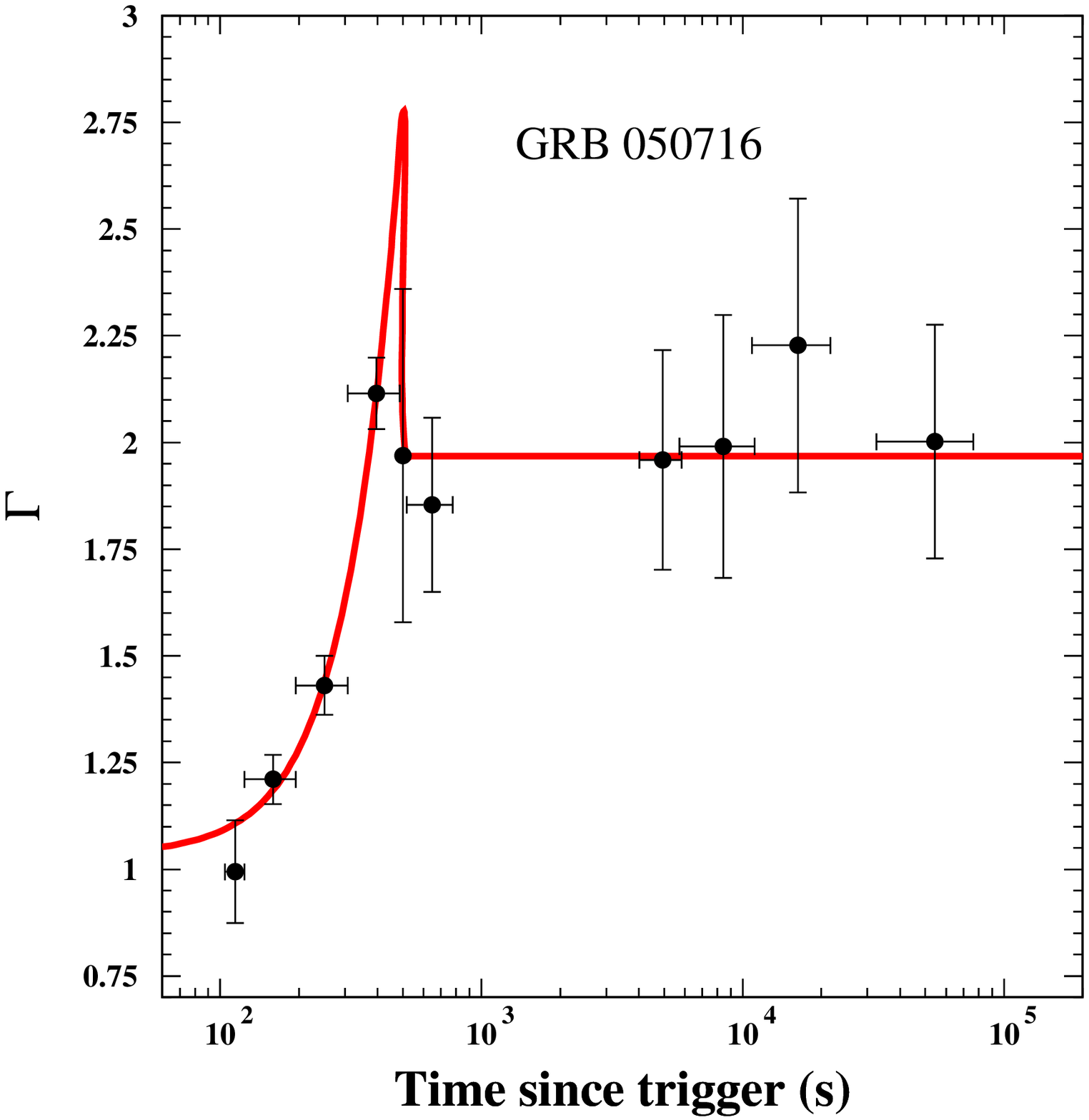,width=8cm}
}}
\caption{
Comparisons between the effective photon spectral
index in the 0.3-10 keV X-ray band as inferred from
observations of GRBs with the Swift XRT,
and the CB model approximate prediction, Eq.~(\ref{Gammat}).
{\bf Top left (a):}  GRB 050814.
{\bf Top right (b):} GRB 050724.
{\bf Bottom left (c):} GRB 050717.
{\bf Bottom right (d):} GRB 050716.
$\Gamma$ values are from {\it http://swift.physics.unlv.edu/xrt.html}.
}
\label{f6}
\end{figure}

In Figs.~\ref{f4} to  \ref{f6} we compare Eq.~(\ref{Gammat}) with the
results for $\Gamma(t)$ for twelve GRBs from the cited web-site
for which the measurements are good. The figures show how the
extracted $\Gamma(t)$ reflects the expected very abrupt transition.
Our simple description of the
observations in terms of three parameters [$t_i$, $\widetilde E/E_p(t_i)$
and $\Delta t$, listed in Table 2], is satisfactory. Also listed in Table 2
are the values of $\Gamma_{\rm SR}$, and the
values of $\alpha\!+\! 1/2$ from our CB-model
fits to the synchrotron-radiation afterglow. 
They are in fair agreement with Eq.~(\ref{alpha}).

\section{Approximate results on more GRBs}

Other authors have analized many more GRBs than we have in this paper.
Zhang et al.~(2007), for instance, confronting the failure of
the high-latitude emission of the FB model to 
explain the rapid softening of the tail of the prompt emission
in sixteen `clean-tail' bright GRBs, proposed an empirical parametrization 
of the X-ray light curve during this phase. Its spectral evolution 
can be rewritten as a time-dependent exponentially cutoff power-law: 
\begin{equation}
F_E\propto \left[{E\over E_c(t)}\right]^{1-\alpha_g}\, 
e^{-E/E_c(t)},\;\;\;\; E_c(t)=E_c(t_i)\,\left({t-t_i\over t_i}\right)^{-k}\, .
\label{Zhang}
\end{equation} 
For  $t\gg t_i$, this is the evolution predicted by the CB model (DD2004), 
provided one identifies $E_c(t)\!=\!E_p(t)$.
Indeed, $E_p(t)\!\approx\!E_p(t_i)$ for $t-t_i\!\ll\!\Delta t_i$,
while for $t-t_i\!\gg\!\Delta t_i$,  
$E_p(t)\!\approx\! E_p(t_i)\, [(t-t_i)/\Delta t_i]^{-k}/2$,
with $k\!=\!2$, see Eqs.~(\ref{Epi},\ref{HRapprox}).
These limiting behaviours may be interpolated by the empirical 
parametrization of Zhang et al.~(2007), in their chosen narrow 
range of $t$, with a constant $k\!\leq \!2$ (they find $1\!\leq\!k\!\leq\!1.6$).  
These authors also discern GRBs without a 
rapid spectral softening during the fast decline. These seem to us
to be cases whose spectral evolution is poorly measured,
or cases, like GRBs 061126 and 061007, whose `fast decline phase' is not 
the end of the prompt emission but the late decline 
of a canonical AG  whose plateau phase ended
 before the beginning of the XRT observations
(DD20007b).

\section{Conclusions}

The spectrum of the $\gamma$-ray peaks and X-ray flares of a GRB or an XRF 
is predicted in the CB model: it is the spectrum of the `glory's light', 
Compton-boosted by the electrons in a CB (DD2004). The time evolution of 
the spectrum traces the voyage of the CB through this `target' light. 
Though the model predicts the spectrum and its evolution at all 
frequencies and times, we have focused on the very rapid decline of the 
flux at the end of a pulse, and the equally swift spectral softening. 
Their understanding is simple: the glory's `target' light is light 
scattered by the circum-burst matter, and its spectrum is exponentially 
cutoff. Its number density, and the flux of a pulse, decrease roughly as 
$1/r^2\!\propto\!1/t^2$. Simultaneously, the target light is becoming more 
radial, so that the characteristic energy of the up-scattered radiation 
also decreases as $1/t^2$. These simple facts, explicitly reflected in 
the predicted `master formula', Eqs.~(\ref{Epi},\ref{GRBXlc}), result in an 
excellent description of the observations.

\begin{figure}[]
\centering
\vbox{
\hbox{
\epsfig{file=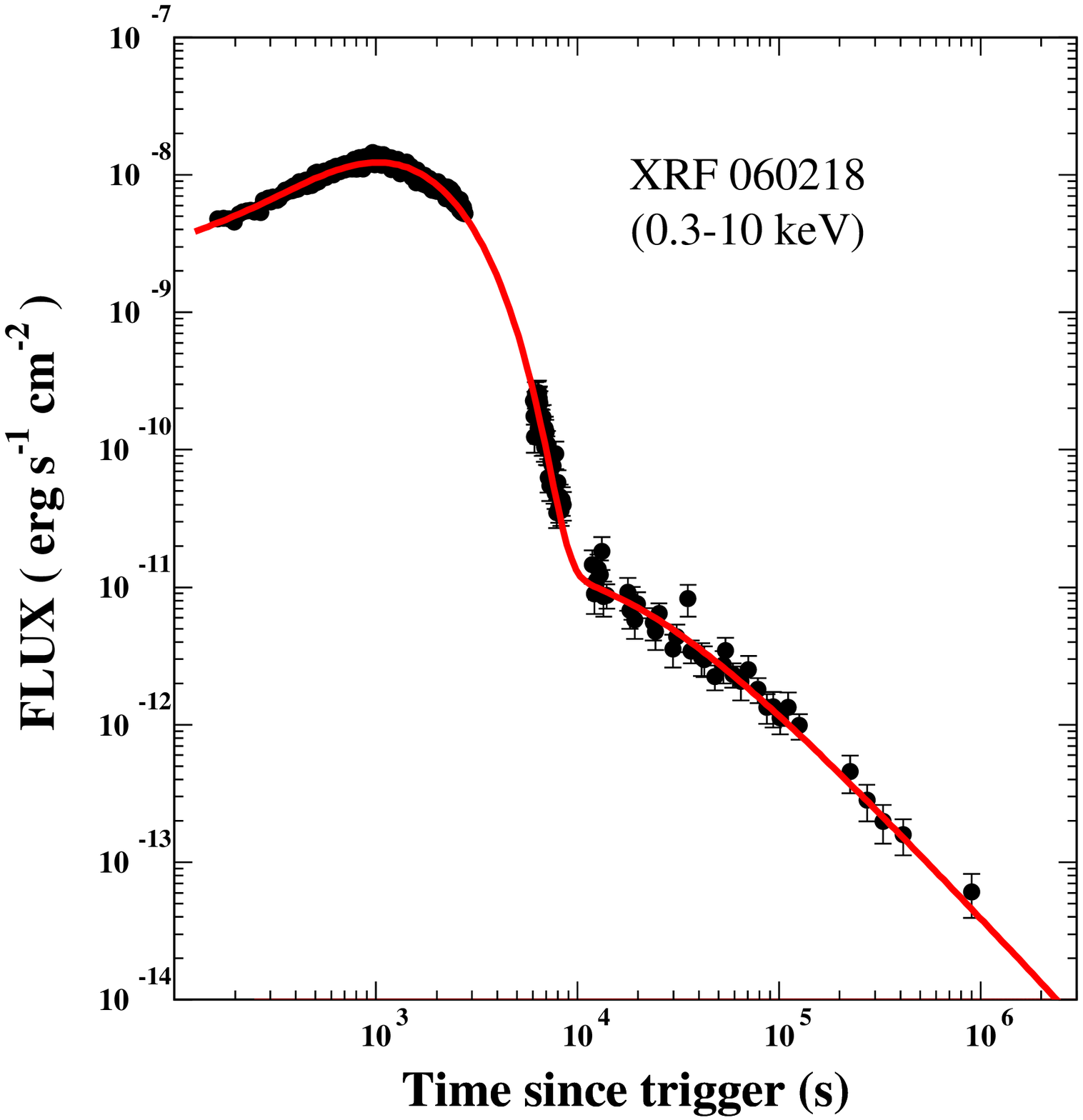,width=8.0cm}
\epsfig{file=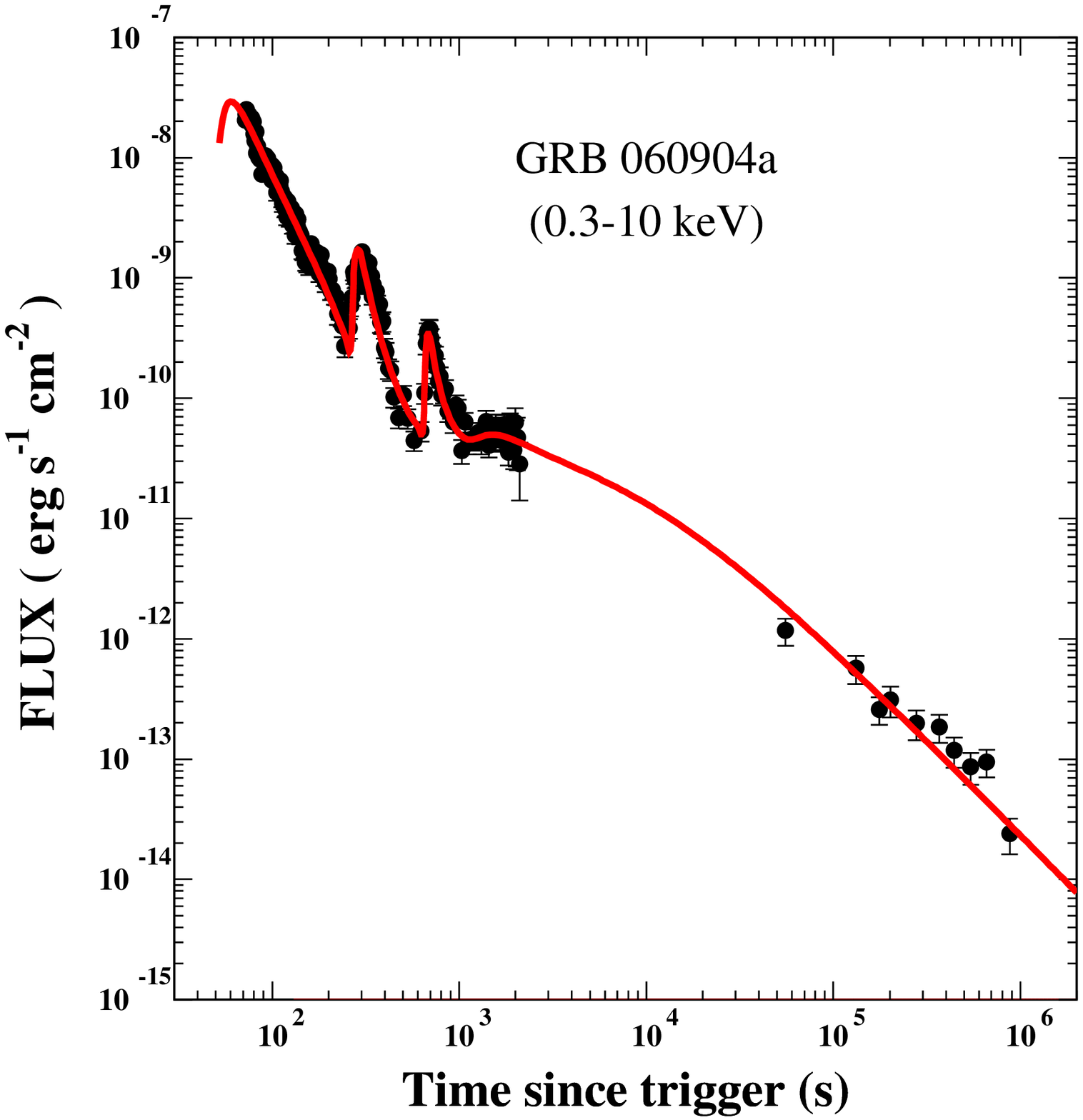,width=8.0cm}
}}

\vbox{
\hbox{
 \epsfig{file=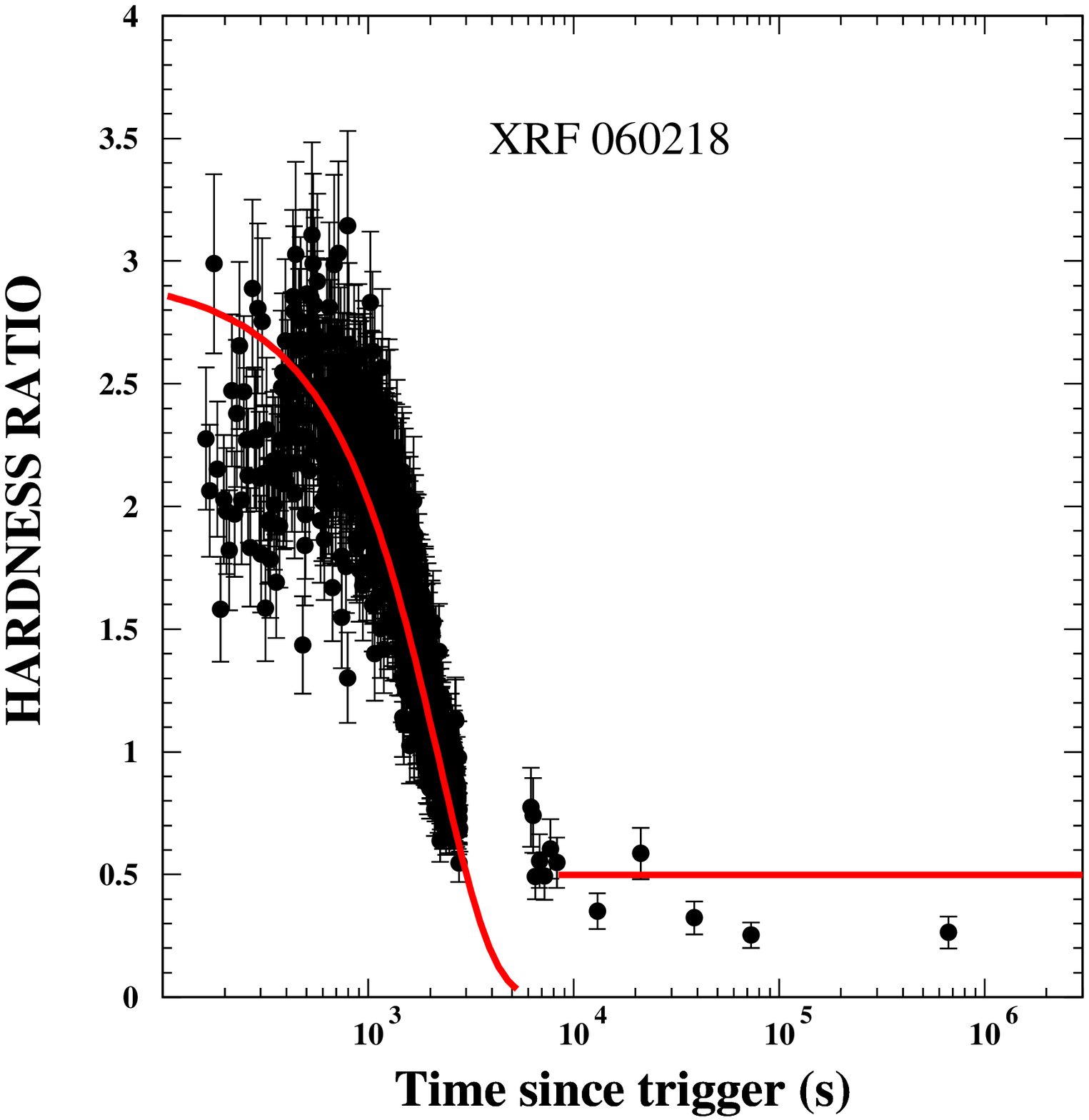,width=8cm}
 \epsfig{file=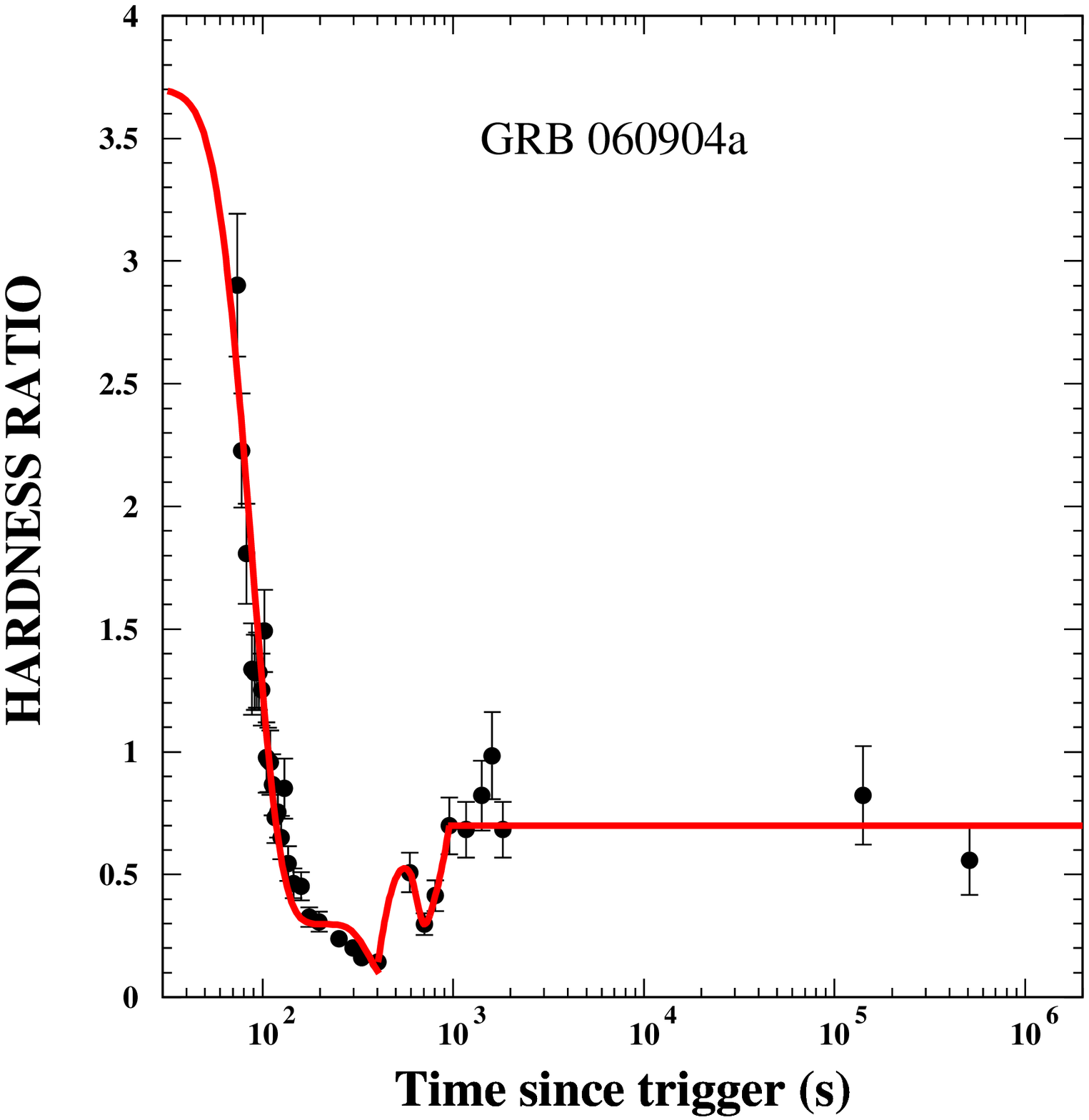,width=8cm}
}}
\caption{
Comparisons between Swift XRT observations 
(Evans et al.~2007) and the CB model predictions.
{\bf Top left (a):} The light curve of XRF 060218.
{\bf Top right (b):} The light curve of GRB 060904a. 
{\bf Bottom left (c):} The hardness ratio of XRF 060218.
{\bf Bottom right (d):} The hardness ratio of GRB 060904a.
}
\label{f1}
\end{figure}

\begin{figure}[]
\centering
\vspace{-1cm}
\vbox{
\hbox{
\epsfig{file=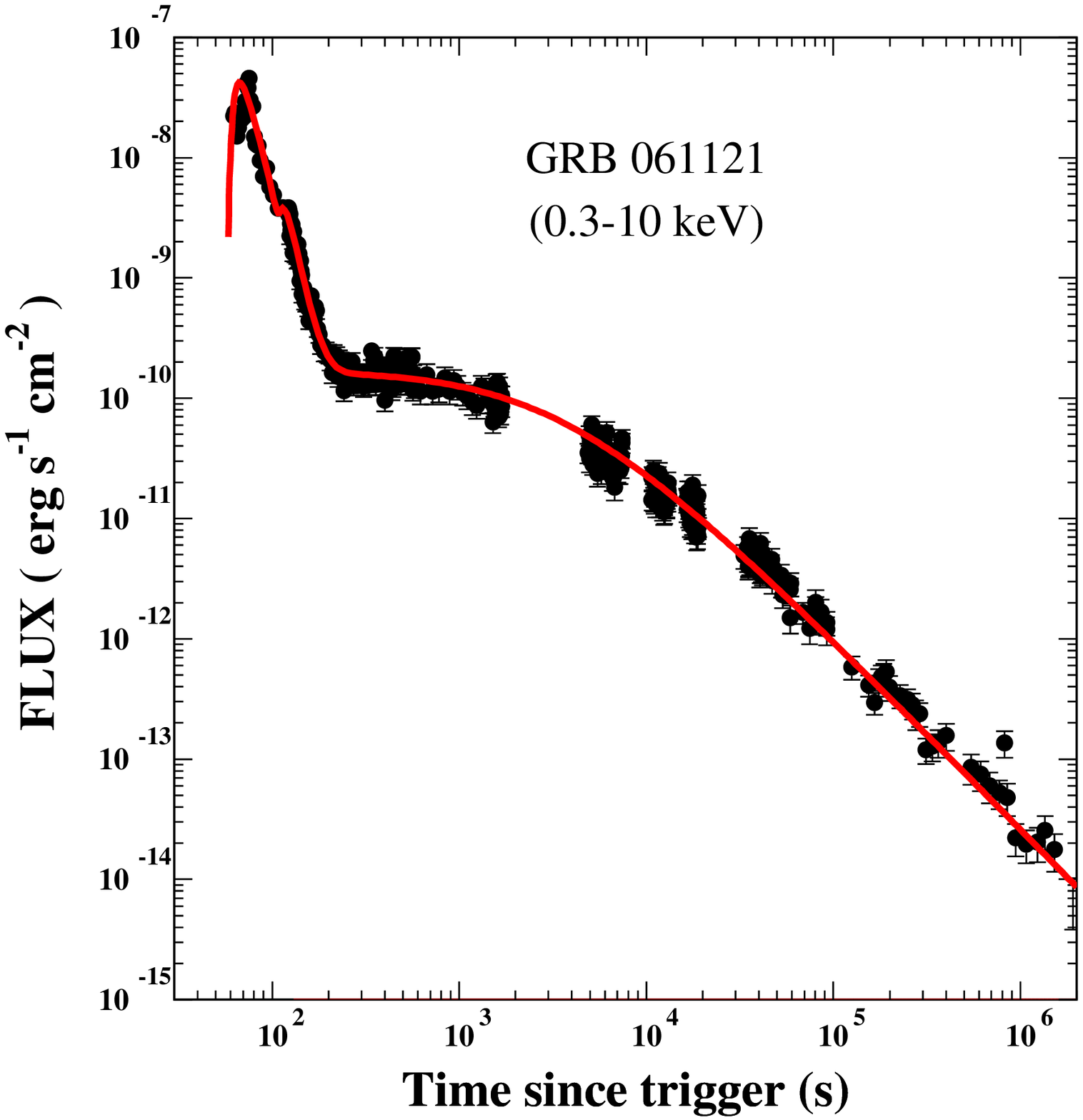,width=8.0cm}
\epsfig{file=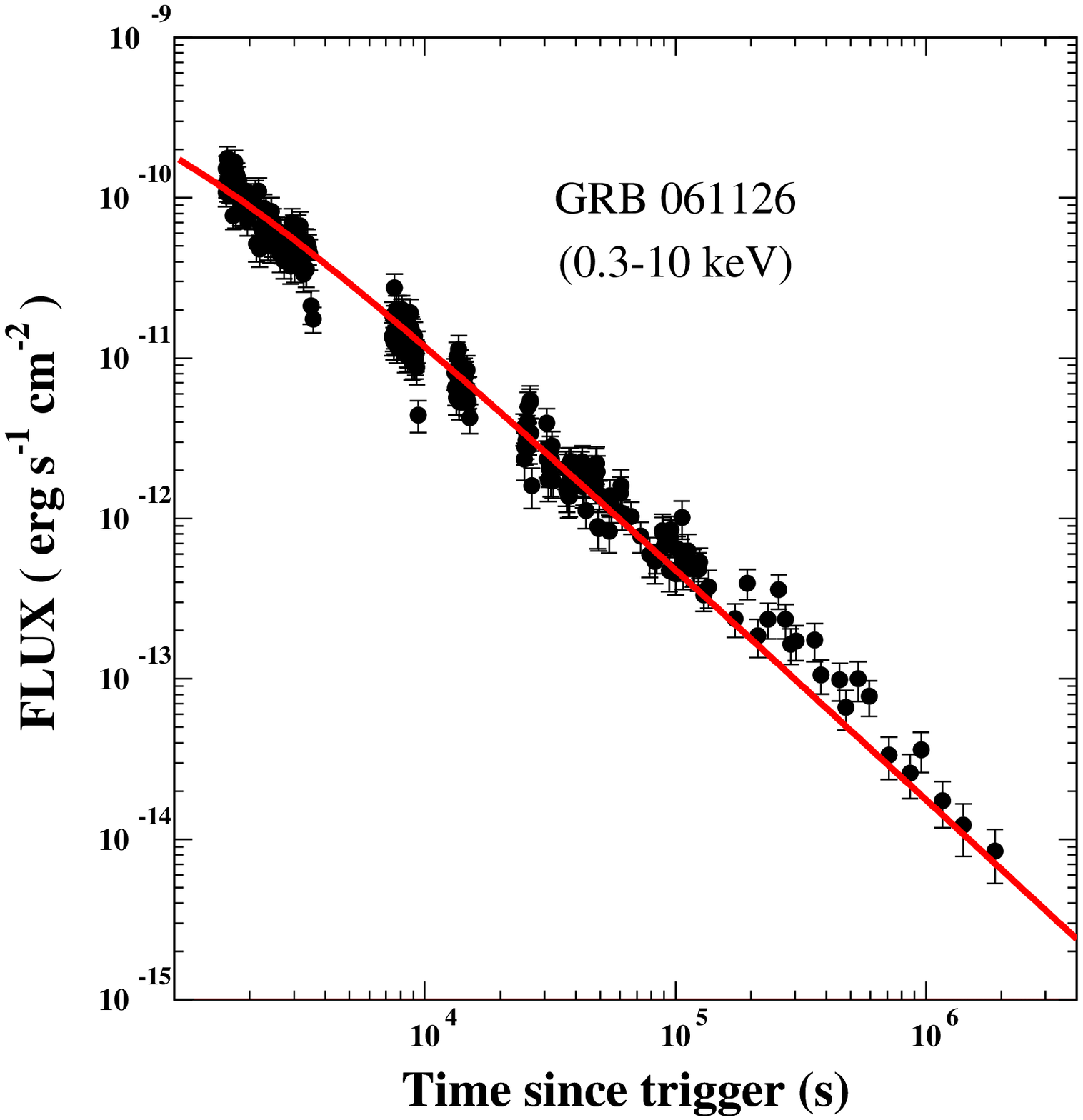,width=8.0cm}
}}
\vbox{
\hbox{
 \epsfig{file=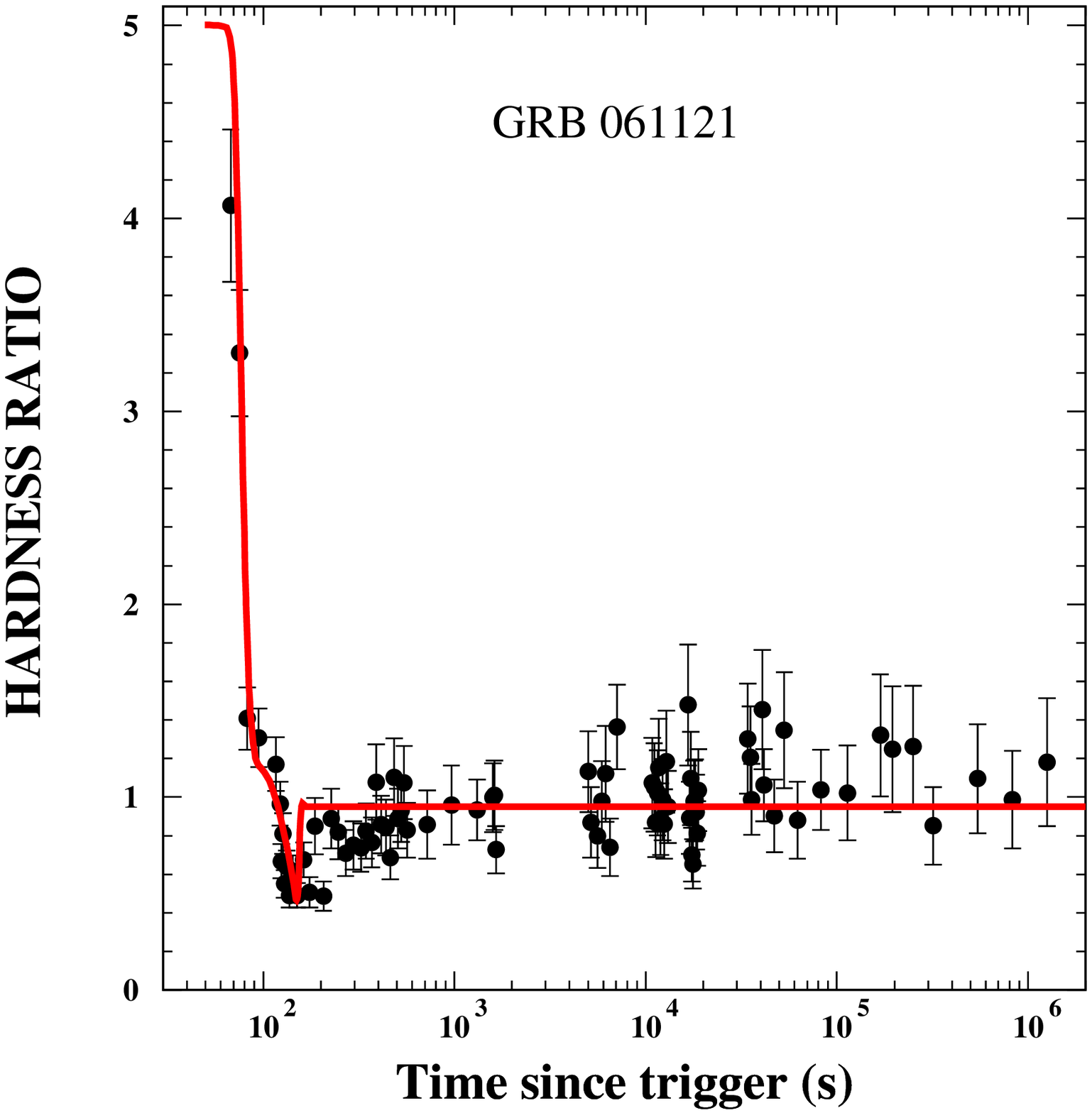,width=8cm}
 \epsfig{file=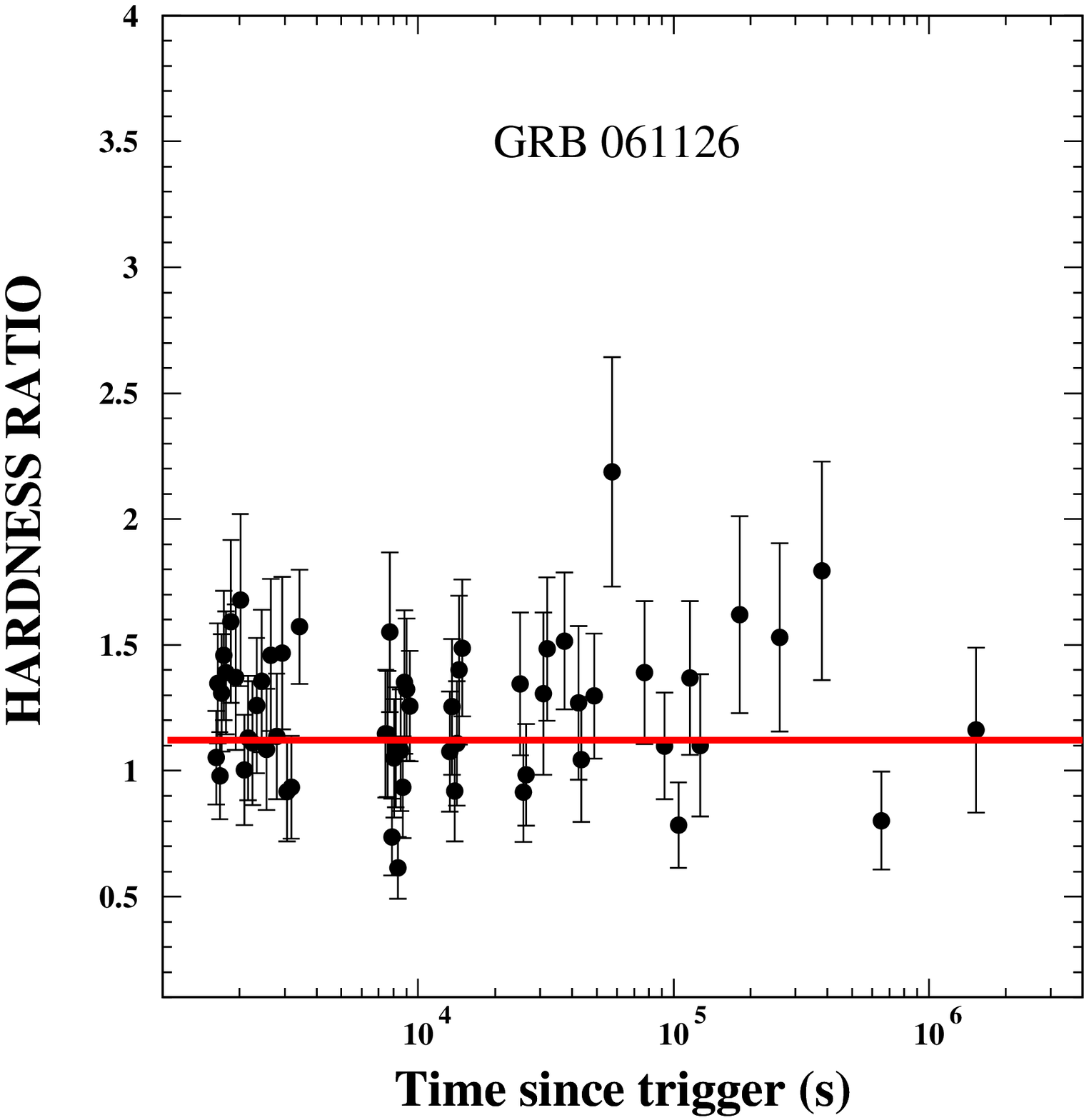,width=8cm}
}}
\caption{
Comparisons between Swift XRT observations
(Evans et al.~2007) and the CB model predictions. 
{\bf Top left (a):} The light curve of GRB 061121.
{\bf Top right (b):} The light curve of GRB 061126. 
{\bf Bottom left (c):} The hardness ratio of GRB 061121.
{\bf Bottom right (d):} The hardness ratio of GRB 061126.
}
\label{f2}
\end{figure}

\begin{figure}[]
\centering
\vbox{
\hbox{
\epsfig{file=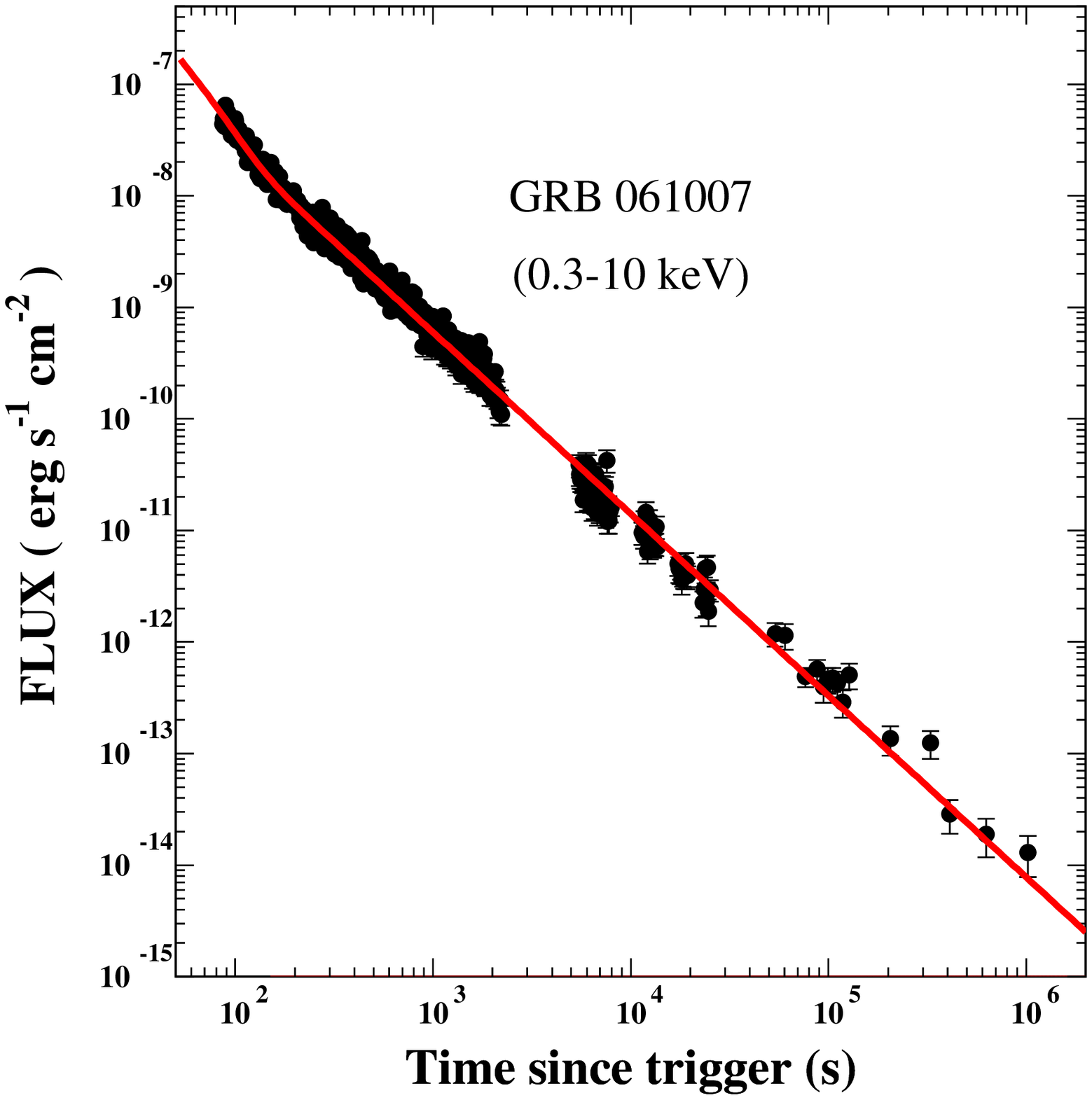,width=8.0cm}
\epsfig{file=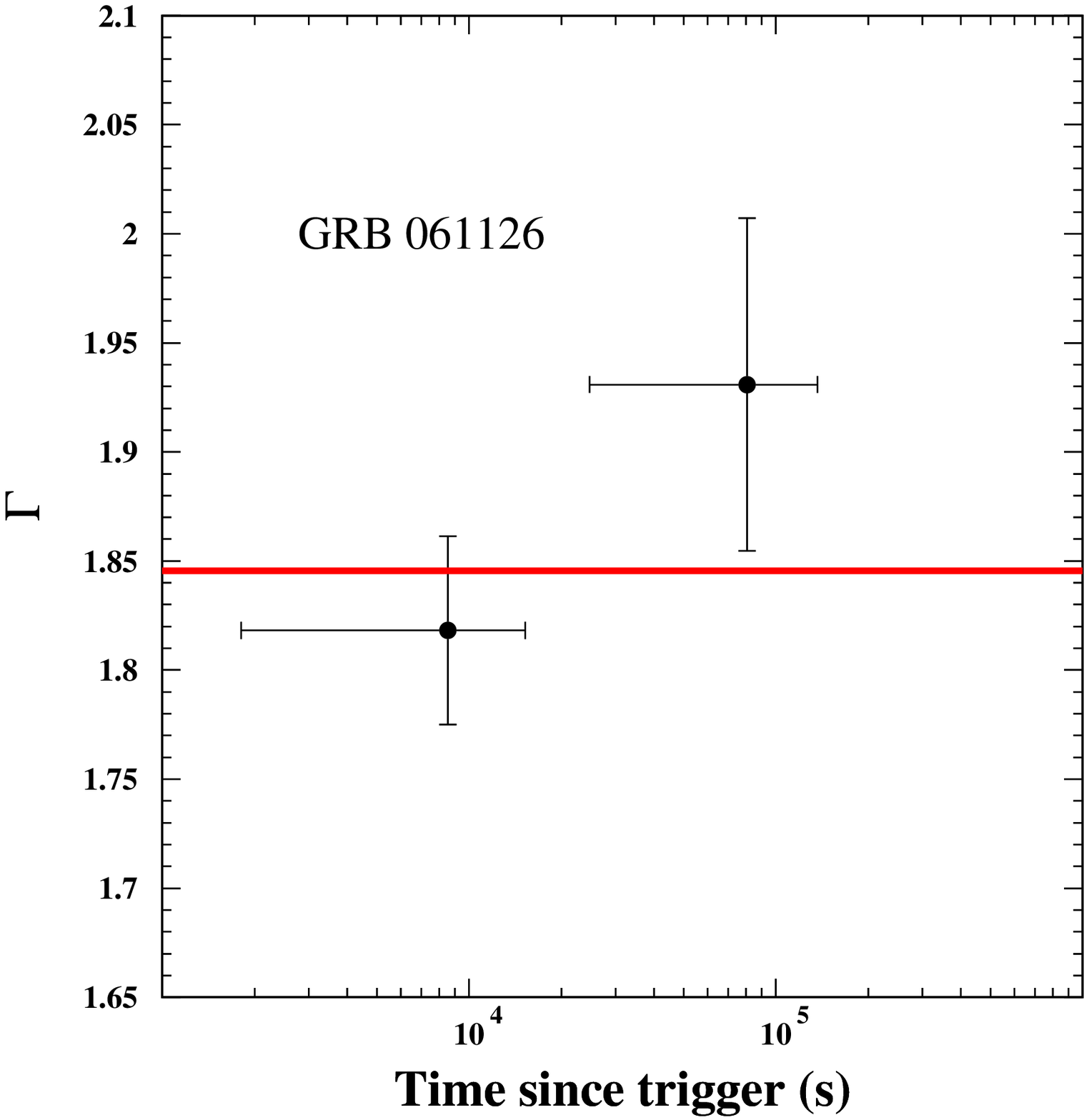,width=8.0cm}
}}

\vbox{
\hbox{
 \epsfig{file=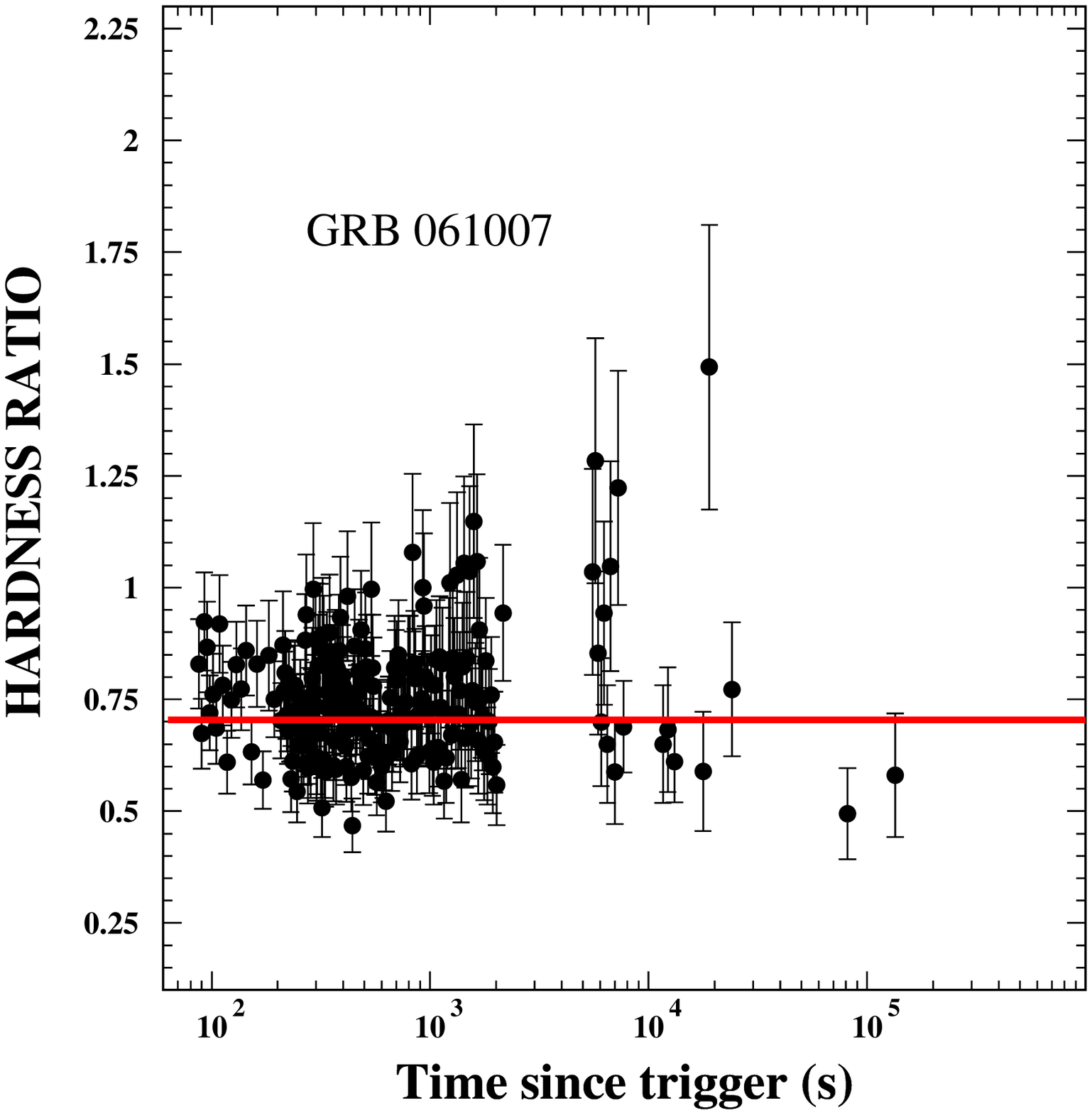,width=8cm}
 \epsfig{file=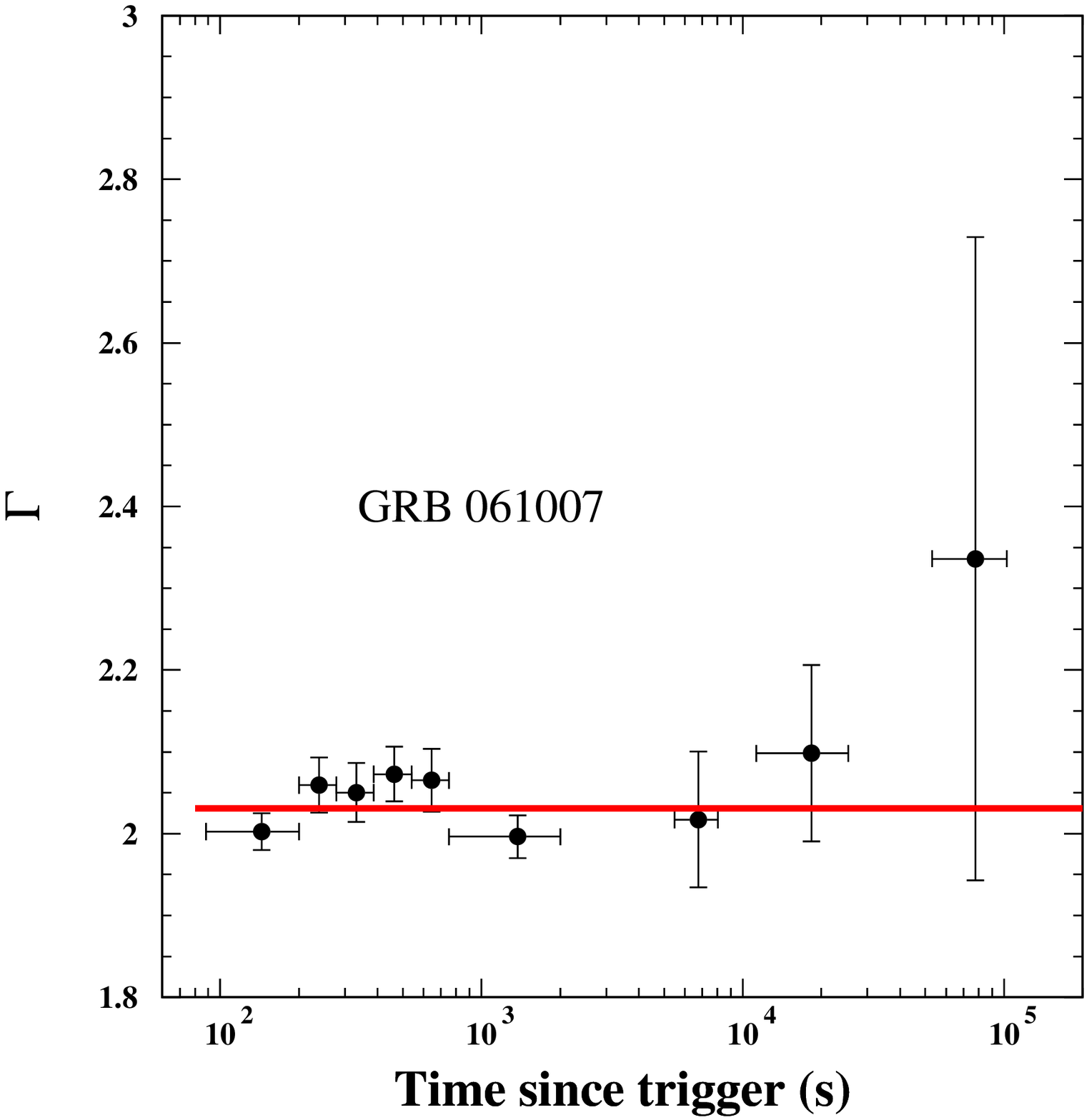,width=8cm}
}}
\caption{
Comparisons between Swift XRT observations (Evans et al.~2007)
and the CB-model predictions.
{\bf Top left (a):} The light curve of GRB 061007.
{\bf Top right lef (b):} The photon spectral index of GRB 061126.
{\bf Bottom left (c):} The hardness ratio of GRB 061007.
{\bf Bottom right (d):} The photon spectral index of GRB 061007.
$\Gamma$ values are from {\it http://swift.physics.unlv.edu/xrt.html}.
}
\label{f3}
\end{figure}

Lacking access to detailed spectral analyses, we have used Swift
data on hardness ratios, uncorrected for X-ray absorption  (Evans et al.~2007), 
as well as the effective spectral index of the unabsorbed 
spectrum reported in {\it http://swift.physics.unlv.edu/xrt.html}.
 We have demonstrated that the spectral time dependences snugly
trace their expected correlation to the corresponding flux variations.
This test of the CB model validates it once again. Yet,
carefully time-resolved absorption corrections would allow even more conclusive 
tests. Time-resolved corrections are 
important because, in the CB model, the line of sight to the hyperluminal CBs 
changes significantly during the long afterglow phase (e.g., DD2004) sweeping
different regions of the host galaxy and  the 
IGM. The changing absorption may  induce flickering of the observed 
X-ray light curve and X-ray spectrum. In fact, the scintillation-like 
behaviour in many X-ray light curves and spectra 
(see Figs.~\ref{f1},\ref{f2},\ref{f3}), if not 
instrumental, may be due to the motion of the CBs in the host 
galaxy. This motion may also 
explain (Dado, Dar \& De R\'ujula, in preparation) the reported time-dependence 
of the equivalent widths of intergalactic absorption systems detected in the 
afterglow of GRB 060206 (Hao et al.~2007, but see also Thone et al.~2007).

At least for GRBs or XRFs with a `canonical' light curve, the 
transition in time from a rapidly falling X-ray decline to a much less steep plateau
--accompanied by the simultaneous and even more pronounced change in the 
spectrum that we have studied-- reflect
one of the most discontinuous transitions seen in astrophysical data.
In the CB model this transition is not attributed to the continued activity of a 
steadily energizing engine, but to the passage from one to another dominant 
radiation mechanism:
inverse Compton scattering versus synchrotron radiation. The transition is so fast
because the late decline of the ICS contribution of Eqs.~(\ref{Epi}, \ref{GRBXlc})
is exponential in time, a consequence of the exponential
cutoff (in energy) of the thin-bremsstrahlung
spectrum of the up-scattered light (DD2004).

\noindent
{\bf Acknowledgment:} The authors would like to thank an
anonymous referee for useful information, comments and suggestions.

\newpage
\begin{deluxetable}{lllllc}
\tablewidth{0pt}
\tablecaption{CB-model afterglow parameters}
\tablehead{
\colhead{Parameter} & \colhead{060218} & \colhead{060904a}& 
\colhead{061121}& \colhead{061126}& \colhead{061007}}
\startdata
$t_1\,[{\rm s}]$         & $-1080$ & 41.08  & 52.48 &  --- & ---\\
$\Delta t_1\,[{\rm s}]$  &  1977 & 16.02  & 12.44 &  --- &--- \\
$\Delta E/E_p(t_1) $     &  0.19 & 0.0452 & 0.061 & --- & --- \\
$t_2\,[{\rm s}]$         & --- & 252.8  & 96.88 &  --- &---\\
$\Delta t_2\,[{\rm s}]$  & --- & 27.75  & 18.80 & --- & ---\\
$\Delta E/E_p(t_2)    $  & ---  & 0.0177 & 0.0014&   --- & --- \\
$t_3\,[{\rm s}]$         & --- & 629.7  &  --- & --- & ---\\
$\Delta t_3\,[{\rm s}]$  & ---  &  44.0  &  ---     &  --- & ---\\
$t_4\,[{\rm s}]$         & ---  & 703.4  &  --- &  --- & ---  \\
$\Delta t_4\,[{\rm s}]$  &  ---     & 747.3  & --- &  --- &--- \\
\hline
$t_0\,[{\rm s}]$         &   183  &  821  & 248&    263 &  40  \\
$\gamma\,\theta$         & 4.28  & 1.25  & 1.42  &  1.12& $\ll 1$ \\
$p$                     & 2.20   & 2.20  & 2.20  &  1.90 & 2.26  \\
\enddata
\label{t1}
\end{deluxetable}

\newpage
\begin{deluxetable}{llllllc}
\tablewidth{0pt}
\tablecaption{Parameters in the description of the photon spectral index
$\Gamma(t)$. The values of $\Gamma_{\rm SR}$ are from the Swift
public data in {\it http://swift.physics.unlv.edu/xrt.html}. 
The values of $\alpha\!+\! 1/2$ are from our CB-model
fits to the synchrotron-radiation afterglow. In the model the two last columns ought
to be equal, see Eq.~(\ref{alpha}). }
\tablehead{
\colhead{GRB} & \colhead{$t_i$[s]} & \colhead{$\Delta$ [s]}&
\colhead{$\widetilde E/E_p(t_i)$}&\colhead{$t_{\rm AG}$ [s]}&
\colhead{$\Gamma_{\rm SR}$} & \colhead{$\alpha+1/2$}}
\startdata
061126  &  ---  & --- &  --- & --- & $1.93\pm 0.12$  & 1.95 \\
061007  &  --- & --- &  ---   & --- & $2.10\pm 0.20$ & 2.13 \\
070129 & 243  & 487 &  0.57 &  1050  & $2.28\pm 0.22 $ & 2.14\\
061222A &  108 & 113  &   1.09 & 195 & $2.15\pm 0.08$ & 2.15\\
061121  &  64 & 5.15 &  0.0035 &161 & $1.99\pm 0.13 $& 2.10 \\
061110A &   3.7  & 219  & 1.056 & 261 & --- & 1.80 \\
060814 & 109 & 295 &  0.68 & 360 & $2.20\pm 0.10$ & 2.16\\
060729 & 131 & 146 & 1.48 & 300 & $2.10\pm 0.10$  &  2.10\\
060510B & 190  &  57  & 0.036 & 460 & $2.14\pm 0.15$ & --- \\
060211A &  0  & 325  &  0.44 & 371& $2.03\pm 0.12$ & 2.04 \\
050814  &  12    & 365  & 0.54 & 361& $1.91\pm 0.09$ & 1.93 \\
050724  &  0     & 154 &  0.23 & 320  & $1.88\pm 0.16$ & 1.86\\
050717  &  0     & 194  & 0.19 & 195 & $1.85\pm 0.12 $ & 1.84\\
050716  &  31    & 96  &   0.037 &496& $1.97\pm 0.11$ & 1.88\\

\enddata
\label{t2}
\end{deluxetable}

\newpage

\newpage

\begin{thebibliography}

\bibitem[1993]{Band1993}
Band, et al.~1993, ApJ, 413, 281

\bibitem[2007]{Band2007}
Burrows, D. N. \& Racusin, J.~2007, arXiv:astro-ph/0702633

\bibitem[2006]{Campana2006}
Campana, S., et al.~2006, Nature, 442, 1008

\bibitem[2002]{DDD2002}
Connaughton, V.~2002, ApJ, 567, 1028

\bibitem[2002]{DDD2002OXAG}
Dado, S., Dar, A. \& De R\'ujula, A.~2002, A\&A, 388, 1079

\bibitem[2003]{DDD2003RADIOAG}
Dado, S., Dar, A. \& De R\'ujula, A.~2003, A\&A, 401, 243

\bibitem[2004]{DDD2004XRF}
Dado, S., Dar, A. \& De R\'ujula, A.~2004, A\&A, 422, 381

\bibitem[2007a]{DDD2007X}
Dado, S., Dar, A. \& De R\'ujula, A.~2007a, arXiv:0706.0880

\bibitem[2007b]{DDD2007XBREAKS}
Dado, S., Dar, A. \& De R\'ujula, A.~2007b, arXiv:0712.1527 

\bibitem[2007]{Dai2007}
Dai, X., et al., 2007, ApJ, 658, 509

\bibitem[1998]{Dar1998}
Dar, A.~1998, ApJ, 500, 93

\bibitem[1999]{DP1999}
Dar, A. \& Plaga, R.~1999, A\&A, 349, 259


\bibitem[2004]{DD2004}
Dar, A. \& De R\'ujula, A.~2004, Physics Reports, 405, 203

\bibitem[2004]{Dermer2004}
Dermer, C. D., 2004, ApJ, 614, 284

\bibitem[2007]{Evans2007}
Evans, P. A.  et al.~2007, A\&A, 469, 379 

\bibitem[2004]{FMN1996}
Fenimore, E. E., Madras, C. D., \& Nayakshin, S., 1996, ApJ, 473, 998

\bibitem[2002]{Giblin2002}
Giblin, T. W. et al.~2002, ApJ, 570, 573 

\bibitem[2007]{Hao2007}
Hao, H., et al.~2007, ApJ, 659, 99

\bibitem[2007]{Kocevski2007}
Kocevski, D. \& Butler, N. 2007, arXiv:0707.4478 

\bibitem[2007]{Kumar}
Kumar, P.,  et al.~2007, MNRAS, 376, L57

\bibitem[2000]{KP}
Kumar, P. \& Panaitescu, A. 2000, ApJ, 541, L51

\bibitem[2006]{Liang2006}
Liang, E. W., et al.~2006, ApJ, 646, 351

\bibitem[2007]{Liang2007}
Liang, E. W., et al.~2007, arXiv:0708.2942

\bibitem[2002]{Mesz2002}
M\'{e}sz\'{a}ros, P.~2002, ARA\&A, 40, 137

\bibitem[2006]{M'esz'aros2006}
M\'{e}sz\'aros, P.~2006, Rept. Prog. Phys.   69,  2259

\bibitem[1998]{M'esz'aros1998}
M\'{e}sz\'{a}ros, P., Rees, M. J. \& Wijers, R. A. M. J.~1998, ApJ, 499, 301


\bibitem[2006]{Mirabal2006}
Mirabal, N., et al. 2006, ApJ, 643, L99

\bibitem[2006]{Nous2006}
Nousek, J., et al.~2006, ApJ, 642, 389

\bibitem[2006]{O'Brien2006}
O'Brien, P. T., et al. 2006, ApJ, 647, 1213 

\bibitem[2007]{Page2007}
Page, K. L., et al. 2007, arXiv:0704.1609
  
\bibitem[2006]{Panaitescu}
Panaitescu, A., et al.~2006, MNRAS, 369, 2059

\bibitem[2007]{Perley2007}
Perley, D. A., et al.~2007, arXiv:astro-ph/0703538

\bibitem[2005]{Pian2005} 
Pian, E., 2006, Nature, 442, 1011

\bibitem[1999]{Piran1999}
Piran, T.~1999, Physics Report, 314, 575

\bibitem[2000]{Piran2000}
Piran, T.~2000, Physics Report, 333, 529

\bibitem[2005]{Piran2005} 
Piran, T.~2005, RvMP, 76, 1143

\bibitem[2002]{Ryde2002}
Ryde, F. \& Svensson, R.~2002, ApJ, 566, 210 

\bibitem[2002]{Rossi2002}
Rossi, E., Lazzati, D. \& Rees, M. J.~2002, MNRAS, 332, 945

\bibitem[1995]{SD1995}
Shaviv, N. J. \& Dar, A.~1995, ApJ, 447, 863

\bibitem[2007]{Thone2007}
Thone, C. C. et al.~2007, arXiv:0708.3448  

\bibitem[2007]{Urata2007}
Urata, Y., et al.~2007, 2007arXiv0707.2826

\bibitem[2007]{Yamazaki2006}
Yamazaki, R., et al.~2006, MNRAS, 369, 311

\bibitem[2007]{Yonetoku2007}
Yonetoku, D., et al.~ 2007, arXiv:0708.3968 

\bibitem[2007]{ZM2007}
Zhang, B.~2007, ChJAA, 7, 1

\bibitem[2002]{ZM2002}
Zhang, B. \& M\'esz\'aros, P.~2002, ApJ, 581, 1236

\bibitem[2004]{ZM2004}
Zhang, B. \& M\'esz\'aros, P.~2004, IJMPA, 19, 2385

\bibitem[2006]{Zhang2006}
Zhang, B., et al.~2006, ApJ, 642, 354

\bibitem[2007]{Z2007}
Zhang, B. B., Liang, E. W. \&  Zhang, B.~2007, ApJ, 666.1002 


\end{thebibliography}
\end{document}